\documentclass[prd,showpacs,superscriptaddress,nofootinbib,twocolumn,twoside]{revtex4}
\usepackage[utf8]{inputenc}
\newcommand{\e}{\epsilon}
\newcommand{\p}{\partial}
\newcommand{\non}{\nonumber}
\newcommand{\bea}{\begin{eqnarray}}
\newcommand{\eea}{\end{eqnarray}}
\newcommand{\bes}{\begin{subequations}}
\newcommand{\ees}{\end{subequations}}
\renewcommand{\L}{{\mathcal{L}}}
\newcommand{\bL}{\bar{{\mathcal{L}}}}
\newcommand{\z}{{\bar z}}

\usepackage{mathrsfs,mathtools,bbold}
\usepackage{amsmath,amssymb}
\usepackage{amsfonts}
\usepackage{verbatim}
\usepackage{graphicx}
\usepackage{mathrsfs,mathtools}
\usepackage{ccaption}
\usepackage[dvipsnames]{xcolor}
\definecolor{SchoolColor}{rgb}{0.6471, 0.1098, 0.1882} 
\usepackage{stmaryrd}
\usepackage{textcomp}
\usepackage[colorlinks=true,linkcolor=black,citecolor=teal,urlcolor=MidnightBlue,filecolor=black]{hyperref}

\newcommand{\unity}{1\hspace{-0.243em}\text{l}}

\renewcommand{\L}{{\mathcal{L}}}

\newcommand{\be}[1]{ \begin{equation}\label{#1} }
\newcommand{\ee}{\end{equation}}

\renewcommand{\phi}{\varphi}

\DeclareMathOperator{\extdm}{d}
\newcommand{\extd}{\extdm \!}

\begin{document}

\title{Wilson Lines and Holographic Entanglement Entropy in Galilean Conformal Field Theories}

\author{Rudranil Basu}
\email{rudrobose@gmail.com}
\affiliation{Saha Institute of Nuclear Physics, Block - AF, Sector - 1, Bidhan nagar, Kolkata 700064, India.}

\author{Max Riegler}
\email{rieglerm@hep.itp.tuwien.ac.at}
\affiliation{Institute for Theoretical Physics, Vienna University of Technology, Wiedner Hauptstrasse 8-10/136, A-1040 Vienna, Austria, Europe}

\date{\today}

\begin{abstract}
In this paper we present in more detail a construction using Wilson lines and the corresponding dual Galilean conformal field theory calculations for analytically determining holographic entanglement entropy for flat space in $2+1$ dimensions first presented in \cite{Bagchi:2014iea}. In addition we show how the construction using Wilson lines can be expanded to flat space higher-spin theories and determine the thermal entropy of (spin-3 charged) flat space cosmologies using this approach.
\end{abstract}


\maketitle

\hypersetup{linkcolor=SchoolColor}

\section{Introduction}

Entanglement entropy as a measure to quantify the amount of entanglement of a quantum system has emerged as a valuable tool in theoretical physics. Exact analytical calculations of entanglement entropy for interacting quantum field theories (QFTs) are, however often hard or even impossible to perform as the computations quickly become intractable with increasing complexity and dimension of the system in question. Two-dimensional conformal field theories (CFTs) with their infinite dimensional symmetries, however, are one of the few examples where exact calculations of entanglement entropy are possible \cite{Holzhey:1994we,Calabrese:2004eu}. Thus QFTs with infinite dimensional symmetries such as Galilean conformal field theories (GCFTs) in two dimensions treated in this paper are a very active playground for various entanglement entropy calculations.\\
The holographic principle \cite{Susskind:1994vu} which states that a theory of (quantum) gravity described by a given spacetime can be equivalently formulated as a QFT living on the boundary of that spacetime has played an important role in the recent surge of interest in entanglement entropy. The most famous example of a realization of this holographic principle is the anti-de Sitter/conformal field theory (AdS/CFT) correspondence which relates Type IIB superstring theory on AdS$_5\otimes$S$^5$ to $\mathcal{N}=4$ supersymmetric Yang-Mills theory on its boundary \cite{Maldacena:1997re}.\\
In 2006 Ryu and Takayanagi conjectured that the computation of entanglement entropy in CFTs is equivalent to computing the area of an extremal codimension-two surface in AdS \cite{Ryu:2006bv}. Thus using this holographic conjecture complex computations of entanglement entropy in higher dimensional CFTs can be replaced by comparatively simple computations of extremal surfaces in AdS. See \cite{Bombelli:1986rw,Srednicki:1993im,Kitaev:2005dm,Ryu:2006ef,Solodukhin:2006xv,Hubeny:2007xt,Nishioka:2009un,Headrick:2010zt,Casini:2011kv,Almheiri:2012rt,Maldacena:2013xja,Marolf:2013dba,Papadodimas:2013wnh,Nozaki:2013wia,Harlow:2014yka,Ecker:2015kna} for a selection of papers on holographic entanglement entropy including precursors to the holographic connection between gravity and quantum entanglement.\\
A particularly interesting aspect of holography is the study of higher-spin symmetries in AdS inspired by the seminal work by Klebanov and Polyakov \cite{Klebanov:2002ja,Mikhailov:2002bp,Sezgin:2002rt}. What makes higher-spin holography so interesting is that it is a weak/weak correspondence \cite{Maldacena:2011jn,Maldacena:2012sf} in contrast to the AdS/CFT correspondence which is a weak/strong correspondence \cite{Gubser:1998bc,Witten:1998qj}. Being a weak/weak correspondence higher-spin holography is thus much more interesting for explicit checks of the holographic principle as calculations are often feasible on both sides of the correspondence. Most of the research in higher-spin holography up until recently was focused on AdS and holographic aspects thereof \cite{Gaberdiel:2010pz}, see \cite{Ammon:2012wc,Gaberdiel:2012uj,Perez:2014pya,Afshar:2014rwa} for selected reviews.\\
For possible future applications to condensed matter physics or checking the generality of the holographic principle it is of interest to try and formulate holographic correspondences which are not asymptotically AdS. Higher-spin gravity, in particular, has turned out to be a very fertile ground to formulate non-AdS holography \cite{Gary:2012ms} including Lobachevsky holography \cite{Afshar:2012nk,Afshar:2012hc}, Lifshitz holography \cite{Gutperle:2013oxa,Gary:2014mca}, de Sitter holography
\cite{Krishnan:2013zya} and flat space holography \cite{Afshar:2013vka,Gonzalez:2013oaa}.\\
An especially nice playground for higher-spin holography is found in $2+1$ dimensions where one can truncate the otherwise infinite tower higher-spin fields, circumvent various no-go results for massless interacting higher-spin theories in flat space \cite{Coleman:1967ad,Aragone:1979hx,Weinberg:1980kq} and formulate gravity in terms of a Chern-Simons gauge theory \cite{Witten:1988hc}. Since notions like geodesics -- extremal codimension-two ``surfaces'' in $2+1$ dimensions-- do not exist for higher-spin gravity, generalizations of the Ryu-Takayanagi proposal are necessary in order to determine entanglement entropy via a holographic description. This has been done in \cite{Ammon:2013hba,deBoer:2013vca} using Wilson lines as the generalization of geodesics.\\
The famous entanglement entropy calculations for 2D CFTs aside there is another interesting class of 2D QFTs, namely Galilean conformal field theories \cite{Bagchi:2009my} which were proposed as duals \cite{Bagchi:2010zz} to $2+1$-dimensional gravity in asymptotically flat spacetimes \cite{Barnich:2006av}. This has sparked considerable interest in a better understanding of flat space holography \cite{Bagchi:2012yk,Bagchi:2012xr,Barnich:2012xq,Barnich:2012aw,Bagchi:2012cy,Bagchi:2013bga,Bagchi:2013lma,Costa:2013vza,Fareghbal:2013ifa,Krishnan:2013tza,Krishnan:2013wta,Bagchi:2013qva,Detournay:2014fva,Barnich:2014kra,Barnich:2014cwa,Riegler:2014bia,Fareghbal:2014qga,Fareghbal:2014kfa}.\\
In \cite{Bagchi:2014iea} entanglement entropy for 2D GCFTs was first calculated both holographically and from a field theory perspective. In this work we build upon this work by extending the holographic setup to accommodate higher-spin symmetries and taking into account GCFTs with finite temperature.

This paper is organized as follows. Section \ref{GCFT} describes in detail the approach and tools necessary to calculate entanglement entropy in Galilean conformal field theories. In Sec. \ref{sec:EEInGCFTs} we then calculate the entanglement entropy for ultrarelativistic field theories exhibiting $\mathfrak{bms}_3$ symmetries at zero and finite temperature using the BMS/GCA equivalence in two dimensions. In Sec. \ref{sec:ThermalEntropy} we describe how to determine the density of states and the corresponding entropy for GCFTs at high temperature. Section \ref{sec:HFSEE} describes a holographic construction which can be used to calculate holographic entanglement entropy for flat spacetimes using Wilson lines and its extension to higher-spin symmetries in flat space. We also determine the holographic entanglement entropy for the null orbifold, global flat space and FSCs explicitly for the spin-2 case. In Sec. \ref{sec:ThermalEntropyFSC} we apply the previously developed formalism and calculate the thermal entropy for spin-2 and spin-3 charged FSCs.

\section{Galilean Conformal Field Theory in Two Dimensions}\label{GCFT}

In this section we explain the basics of GCFTs in 2D and then use this knowledge to compute entanglement entropy  first for a GCFT at zero temperature and then for a GCFT at finite temperature. More details on the mathematical tools used in the following sections can be found in \cite{Bagchi:2009pe}.

\subsection{Kinematics}

First let us consider a system in two dimensions invariant under Galilei transformations. If we demand in addition scaling invariance of the system generated by a dilatation operator, one obtains the finite dimensional Galilean conformal group. This is a six-dimensional group generated by a Hamiltonian (${H \sim \p_t}$), momentum (${P \sim \p_x}$), Galilean boost (${G \sim t \p_x}$), dilatation (${D \sim t \p _t + x \p_x}$) and two transformations (${K \sim 2tx \p_x + t^2 \p_t}$) and (${B \sim t^2 \p_x}$) which can be thought of as the Galilean conformal analogues of the spatial and temporal components of relativistic special conformal transformations. It is convenient to choose a more compact notation for these generators given by
\bea \label{gcavf}
L_n = - t^{n+1} \p_t - (n+1) t^n x \p_x , \quad M_n = t^{n+1} \p_x,
\eea
with $n= 0, \pm 1$. As vector fields, the Lie brackets of the generators form the following Lie algebra also known as the finite Galilean conformal algebra
\bea \label{gca}
\left[L_n, L_m\right] &=& (n-m) L_{n+m},  \non \\
\left[L_n, M_m\right] &=& (n-m) M_{n+m},  \non \\
\left[M_n, M_m\right] &=& 0.
\eea
This Lie algebra \eqref{gca} closes even if one includes an infinite set of vector fields by letting $n$ in \eqref{gcavf} run over all integers. We will denote this as Galilean conformal algebra  (GCA) from now on. However closure of this infinite extension is not a great surprise if one thinks about the GCA as a nonrelativistic limit of the relativistic conformal algebra in two dimensions. The relativistic conformal group is generated by the Witt algebra whose Lie brackets are given by
\bea \label{witt}
\left[\L_n, \L_m\right] &=& (n-m) \L_{n+m}, \nonumber\\
\left[\bL_n, \bL_m\right] &=& (n-m) \bL_{n+m}, \nonumber\\
\left[\L_n, \bL_m\right] &=& 0,
\eea
 of the set of vector fields
\bea \label{witt_vf}
\L_n = z^{n+1} \p_z, \quad \bL_n = \z^{n+1} \p_\z,
\eea
on the complex plane. The limiting process which yields the GCA is an \.In\"on\"u--Wigner contraction and can be realized via 
\bea \label{nr}
L_n = \L_n + \bL_n  \quad\mbox{ and }\quad M_n = \e \left( \L_n - \bL_n\right),
\eea
for some real parameter $ \e$. In the limit $ \e \rightarrow 0$, one recovers \eqref{gca} by using the relations \eqref{witt}. The contraction \eqref{nr} is essentially a transformation of the structure constants of the Witt algebra. However, there is an associated spacetime interpretation of the above contraction which is compatible with the vector field representations \eqref{gcavf} and \eqref{witt_vf} of the generators. From the spacetime point of view, Lorentz invariance should be broken when going to a Galilean regime, which can be most easily achieved by unequal scaling of the spatial and the temporal coordinate. Expressing the coordinates of \eqref{witt_vf} in Lorentzian signature as $z= t+ x,\,\z = t-x$ one can introduce the following scaling
\bea \label{scaling}
 t \rightarrow  t, ~~ x \rightarrow \e x.
\eea
The linear combinations \eqref{nr} in the limit $ \e \rightarrow 0$ then again yield the vector fields \eqref{gcavf}.\\
The analysis above shows how a theory with GCA symmetry can be understood as a particular limit of a 2D CFT. However, it should be appreciated as well that interchanging the notions of space and time in a two-dimensional system should leave many structural features unaffected. Building upon this idea, another interesting scaling can be applied to the CFT given by
\bea \label{scaling2}
 t \rightarrow \e t, ~~ x \rightarrow x.
\eea
In contrast to \eqref{scaling}, this is the ultrarelativistic limit of the relativistic parent CFT which is actually the relevant limit in the context of flat space holography.

\subsection{Quantization and Highest Weight Representation}

Similar to the Virasoro algebra which is a centrally extended version of the Witt algebra \eqref{witt} the GCA also admits central extensions of the following form:
\bea \label{gcaex}
\left[L_n, L_m\right] &=& (n-m) L_{n+m}  + \dfrac{c_L}{12}(n^3 - n) \delta_{m+n,0},\non \\
\left[L_n, M_m\right] &=& (n-m) M_{n+m} + \dfrac{c_M}{12}(n^3 - n) \delta_{m+n,0},\non \\
\left[M_n, M_m\right] &=& 0.
\eea
As mentioned in the previous subsection in 2D, Eq. \eqref{gcaex} can be obtained by two different limits from two copies of the Virasoro algebra. We want to stress at this point that we are mainly interested in the algebra \eqref{gcaex} and its highest weight representations which ware independent of any limit. The reason for this is that these two limits only differ by a trivial exchange of time and space $t\leftrightarrow x$.\\
In order to compute the entanglement entropy of a GCFT we first need to introduce the notion of a highest weight representation. This representation is fixed by the highest weight state $|h_L, h_M \rangle$ defined as:
\bea \label{hws}
&& L_0 |h_L, h_M \rangle = h_L |h_L, h_M \rangle, \non \\
&& M_0 |h_L, h_M \rangle = h_M |h_L, h_M \rangle, \non \\
&& L_n |h_L, h_M \rangle = M_n |h_L, h_M \rangle = 0 ~\mbox{ for}\, n>0 .
\eea
Repeated application of $L_{-n}$ and $M_{-n}$ for $n>0$ creates new states in this representation. As in a CFT, one can introduce operators corresponding to each of these states. The GCA primaries are local operators $\Phi_{h_L, h_M} (x,t)$ which map the vacuum state to the highest weight state
	\begin{equation} 
		\Phi_{h_L, h_M} (0,0) | 0\rangle  = |h_L, h_M \rangle .
	\end{equation}
The transformation properties of the primaries under the Galilean conformal transformations can be easily derived from first principles by using ${[L_0,\Phi_{h_L, h_M} (0,0)] = h_L\Phi_{h_L, h_M} (0,0)}$ and similarly for $M_0$. Abbreviating ${\Phi_{h_L, h_M} (x,t)\equiv\Phi}$ one thus obtains
\bes
\bea
\delta_{L_n} \Phi&=& \Big[t^{n+1} \p _t + (n+1) t^n x \p_x, \non \\ 
                  &&  + (n+1)t^{n-2} (h_L t - n\,h_M x) \Big]\Phi, \\
\delta_{M_n} \Phi &=&  \left[-t^{n+1} \p _x + (n+1) t^n h_M \right]\Phi.
\eea
\ees
These relations can be more conveniently encoded in a pair of fields which one interprets as Galilean energy-momentum tensors \cite{Bagchi:2010vw}
\bes\label{em_def}
\bea
 T_{(1)} (x,t) &=& \sum _{n} t^{-n-2} \left[ L_n+ (n+2)\frac{x}{t}M_n \right],~\\
 T_{(2)} (x,t) &=& \sum _{n} t^{-n-2} M_n,
\eea
\ees
which is in analogy to the mode expansion of the energy-momentum tensor in terms of Virasoro generators in a CFT.\\
With these definitions at hand, one can now determine Galilean conformal Ward identities. For the purpose of this paper we are primarily interested in Ward identities involving two primary fields ${\Phi_{h^{(i)}_L, h^{(i)}_M} (x_i, t_i)\equiv\Phi^{(i)}}$, with $i=1,2$. One can, for example, determine the Ward identities for $T_{(2)}(x,t)$ via
\begin{align} \label{ward2}
&\langle T_{(2)}\Phi^{(1)} \Phi^{(2)}\rangle=\sum _{n=-1}^{\infty} t^{-n-2} \langle 0| \left[M_n, \Phi^{(1)} \Phi^{(2)}\right]  |0 \rangle\nonumber\\  
&=\sum_{i=1,2}\sum _{n =-1}^{\infty} t^{-n-2}\left[ -t_i ^{n+1} \partial _{x_i} +(n+1) h^{(i)}_M \, t_i ^n \right]\langle \Phi^{(1)} \Phi^{(2)} \rangle  \nonumber \\
&= \sum_{i=1,2} \left( \frac{h^{(i)}_M}{(t-t_i)^2} - \frac{1}{t-t_i} \partial _{x_i}  \right) \langle \Phi^{(1)} \Phi^{(2)}\rangle.
\end{align}
The Ward identity involving $T_{(1)}(x,t)$ can be derived along similar lines and is given by
\begin{align} \label{ward1}
&\langle T_{(1)}\Phi^{(1)} \Phi^{(2)}\rangle=\sum_{i=1,2} \Big [ \frac{1}{(t-t_i)} \partial_{t_i}+ 2 h^{(i)}_M \, \frac{x-x_i}{(t-t_i)^3} \nonumber\\
&+ \frac{1}{(t-t_i)^2} \left( h^{(i)}_L - (x-x_i) \partial_{x_i}\right) \Big]\langle \Phi^{(1)} \Phi^{(2)}\rangle.    
\end{align}
In order to proceed one needs to know the exact form of the two-point correlation function for GCFT primaries which is given up to some normalization constant by
\begin{equation}\label{2_pt}
\langle \Phi^{(1)} \Phi^{(2)}\rangle \sim \delta_{h^{(1)}_L \, h^{(2)}_L } \delta_{h^{(1)}_M \, h^{(2)}_M } \, t_{12}^{-2 h^{(1)}_L} \exp \left(-2 h^{(1)}_M \frac{x_{12}}{t_{12}} \right).
\end{equation}
Using this expression for the two-point function and inserting this into \eqref{ward2} and \eqref{ward1}, one arrives at
\begin{subequations} \label{ward_exp}
	\begin{align} 
 \langle T_{(1)}\Phi^{(1)} \Phi^{(2)}\rangle =& \left(\frac{t_{12}}{t_{{\prime}1} t_{{\prime}2}}\right)^2 t_{12}^{-2h_L} \exp \left(-2 h_M \frac{x_{12}}{t_{12}} \right)\times\nonumber\\ &\times\left[ h_L - 2 h_M \left(\frac{x_{12}}{t_{12}}-\frac{x_{{\prime}1}}{t_{{\prime}1}}-\frac{x_{{\prime}2}}{t_{{\prime}2}}\right)\right], \\
\langle T_{(2)}\Phi^{(1)} \Phi^{(2)}\rangle =& h_M \left(\frac{t_{12}}{t_{{\prime}1} t_{{\prime}2}}\right)^2 t_{12}^{-2h_L} e^{- 2 h_M \frac{x_{12}}{t_{12}}},
	\end{align}
\end{subequations}
where $t_{\prime 1} = t^{\prime}- t_1, t_{12}= t_1-t_2,\ldots$ and we have inserted the energy-momentum tensor at $(x^{\prime},t^{\prime})$.\\
Before ending this section we want to point out a crucial difference in the quantization used in GCFTs and the one generally used in CFTs. In relativistic CFTs radial quantization is a standard procedure of choosing space and time in a Euclidean setup. Starting from an infinitely extended cylinder, one can map this cylinder to the complex plane where in polar coordinates the (Euclidean) time coordinate corresponds to the radial coordinate and equal time spatial foliations correspond to concentric circles. In GCFTs, this is quite different. Since the temporal and spatial coordinates scale differently, the notion of radial chart gives way to Cartesian time and space coordinates. The phrase "equal time" in the context of canonical quantum theory would now mean noncompact spatial dimension $\sim \mathbb{R}$.

\subsection{Transformation Properties of the Energy Momentum Tensor}

In the last subsection we showed how to determine the Galilean conformal Ward identities. As in the more familiar relativistic setup these are intimately related with the transformation properties of primary fields. In this subsection we show how to determine the transformation rules of the components of the energy-momentum tensor, $T_{(1)}$ and $T_{(2)}$ under Galilean conformal transformations.\\ 
First let us note that arbitrary diffeomorphisms ${(t,x) \rightarrow (t^{\prime}, x^{\prime})}$ are not compatible with Galilean conformal transformations. The form of the most general transformation of the coordinates $(t,x)$ which are compatible with Galilean conformal transformations can be determined via \eqref{gcavf}. These coordinate transformations are given by \cite{Witten:1988hc}
\bea \label{cr}
t = f(t^{\prime}), ~~ \mbox{ and }~~ x=\frac{df (t^{\prime})}{d t^{\prime}} x^{\prime} + g(t^{\prime}),
\eea
where $f, g$ are arbitrary functions of $t'$. These transformations can be seen as the Galilean conformal analogues of the holomorphic and antiholomorphic transformations generated by the Virasoro vector fields in a relativistic CFT. From \eqref{cr} one can also straightforwardly determine the relations
	\begin{equation}
		\frac{\p t}{\p t^{\prime}} = \frac{\p x}{\p x^{\prime}}, 
		\quad \frac{\p t}{\p x^{\prime}} = 0
	\end{equation}
whose structure resembles the Cauchy-Riemann equations.\\
Similar to a relativistic CFT one can determine the transformation properties of $T_{(1)}$ and $T_{(2)}$ under Galilean conformal transformations by integrating the infinitesimal transformation relations which are determined by the two-point correlators \cite{Bagchi:2015wna} of the energy-momentum tensor with itself. A very useful cross-check of the results obtained this way is given by taking the Galilean limits of the corresponding CFT results. Thus, the \.In\"on\"u--Wigner contraction used in \eqref{nr} can also be used to perform a limit of the CFT transformation rules of the energy-momentum tensor. In order to be compatible with the definition of the components of the GCA energy-momentum tensor \eqref{em_def} one has to take the following linear combinations of the holomorphic and antiholomorphic components of the relativistic CFT energy-momentum tensor:
\begin{gather}
T_{(1)}(t,x) = \lim_{\e \rightarrow 0} \left(T(z) + \bar{T}(\bar{z})\right),\\
T_{(2)}(t,x) = \lim_{\e \rightarrow 0} \e\left(T(z) - \bar{T}(\bar{z})\right),
\end{gather}
where $z=t+\epsilon x$ and $\bar{z}=t-\epsilon x$ along with the scaling rules \eqref{scaling}. This yields
	\begin{subequations}\label{em_trans}
		\begin{align}
			T_{(1)}(t',x') \rightarrow& \left(\frac{\extd t}{\extd t'}\right)^2\,T_{(1)}(t,x)+2 \left(\frac{\extd t}{\extd t'}\right)\left(\frac{\extd x}{\extd t'}\right)\, T_{(2)}(t,x)\nonumber\\
			&+\frac{c_L}{12}\{t,t'\}+\frac{c_M}{12}\left(\frac{\extd t}{\extd t'}\right)^{-1}\llbracket (t,x),t'\rrbracket,\\
			T_{(2)}(t',x') \rightarrow & \left(\frac{\extd t}{\extd t'}\right)^2\,T_{(2)}(t,x)+\frac{c_M}{12}\{t,t'\}.
		\end{align}
	\end{subequations}
where $\{,\}$ denotes the Schwarzian derivative given by
	\begin{equation}
		\{t,t'\}=\left[\left(\frac{\extd^3t}{\extd t'^3}\right)-\frac{3}{2}\left(\frac{\extd^2t}{\extd t'^2}\right)^2\left(\frac{\extd t}{\extd t'}\right)^{-1}\right]\left(\frac{\extd t}{\extd t'}\right)^{-1},
	\end{equation}
and $\llbracket,\rrbracket$ denotes the corresponding Galilean conformal equivalent (GCA Schwarzian) thereof which we defined by
	\begin{equation}
		\llbracket (t,x),t'\rrbracket:=\textrm{\textlquill}(t,x),t'\textrm{\textrquill}-\left(\frac{\extd x}{\extd t'}\right)\{t,t'\},
	\end{equation}
where
	\begin{align}
		\textrm{\textlquill}(t,x),t'\textrm{\textrquill}=&\left(\frac{\extd^3x}{\extd t'^3}\right)+3\left[\frac{1}{2}\left(\frac{\extd^2t}{\extd t'^2}\right)^2\left(\frac{\extd x}{\extd t'}\right)\left(\frac{\extd t}{\extd t'}\right)^{-1}\right.\nonumber\\
		&\left.-\left(\frac{\extd ^2t}{\extd t'^2}\right)\left(\frac{\extd^2x}{\extd t'^2}\right)\right]\left(\frac{\extd t}{\extd t'}\right)^{-1}.
	\end{align}
The exact form of $\{,\}$ and $\llbracket,\rrbracket$ and their appearance together with the central charges $c_L$ and $c_M$ can also be understood as follows. As in a relativistic CFT, the quantum corrections that the classical transformation law of the energy-momentum tensor obtains should vanish for $c_L=0$, $c_M=0$ and global ISL$(2,\mathbb{R})$ transformations. Thus, it is clear that the quantum corrections have to depend on $c_L$ and $c_M$. In addition, whatever $\{,\}$ and $\llbracket,\rrbracket$  are they have to be compatible with the group composition law that two successive transformations $(t,x)\mapsto(t',x')\mapsto(t'',x'')$ yield the same result as mapping $(t,x)\mapsto(t'',x'')$ directly. This together with the invariance under global ISL$(2,\mathbb{R})$ transformations, i.e.,
	\begin{align}
		\{f[t],t\}=&\left\{\frac{af[t]+b}{cf[t]+d},t\right\},\nonumber\\
		\left\llbracket (f[t],g[x]),t\right\rrbracket=&\left\llbracket \left(\frac{af[t]+b}{cf[t]+d},\frac{g[x]}{(d+cf[t])^2}\right),t\right\rrbracket,
	\end{align}
where $a,b,c,d$ are some constants with $ad-bc=1$, determines the form of $\{,\}$ and $\llbracket,\rrbracket$ uniquely.\\
As an addendum we want to note that in the same sense that the Schwarzian derivative measures the degree to which a function fails to be a fractional linear transformation i.e.
	\begin{equation}
		\{t,t'\}=0\quad\Leftrightarrow\quad t(t')=\frac{at'+b}{ct'+d},
	\end{equation}
the GCA Schwarzian also measures the degree to which two functions fail to be Galilean conformal i.e.
	\begin{gather}
		\llbracket(t,x),t'\rrbracket =0\nonumber\\
		\Updownarrow\nonumber\\
		t(t')=\frac{at'+b}{ct'+d},\quad x(t',x')=\frac{x'}{(d+ct')^2}.
	\end{gather}

\section{Entanglement Entropy in GCFTs}\label{sec:EEInGCFTs}

\begin{figure}[h]
		\centering
		\includegraphics[width=0.5\textwidth]{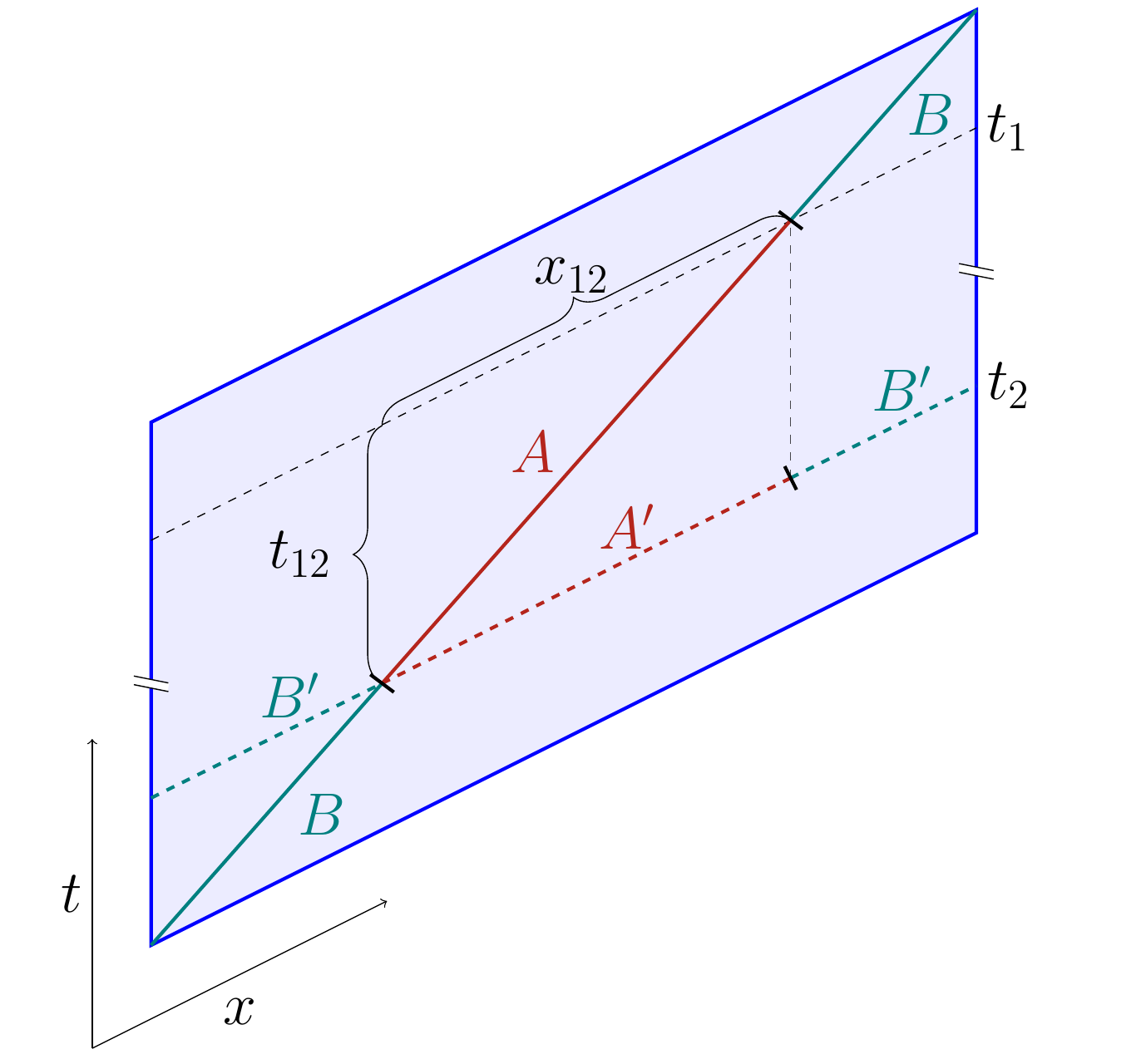}
		\caption{Boosted ({\color{BrickRed}$A$}, {\color{teal}$B$}) and equal time ({\color{BrickRed}$A'$}, {\color{teal}$B'$}) entangled intervals used to determine entanglement entropy in GCFTs.}
		\label{fig:GCFTPicture}
	\end{figure}

In this section we calculate the entanglement entropy of a one-dimensional subsystem in a 2D GCFT in close analogy to \cite{Calabrese:2004eu}. We consider a subsystem $A$ given by a line connecting the points $(x_1, t_1)$ and $(x_2, t_2)$ and its complement, which we call $B$, as shown in Fig. \ref{fig:GCFTPicture}.\\
The motivation for considering such an interval lies in the nonrelativistic nature of a GCFT. In a Lorentz invariant theory observables are not sensitive to a certain choice of frame. Hence, assuming one quantized some Lorentz invariant theory with respect to some time coordinate $t$, one can compute entanglement entropy simply on a $t=0$ slice. A GCFT, however, is not a Lorentz invariant theory and thus observables are sensitive to a choice of frame. Thus, in order to determine how the entanglement entropy in a GCFT depends on the choice of frame, we use a (Galilean) boosted interval ($A$) bounded by the points $(x_1, t_1)$ and $(x_2, t_2)$ instead of an equal time interval ($A'$) which would be bounded by $(x_1, t_2)$ and $(x_2, t_2)$ as depicted in Figure \ref{fig:GCFTPicture}.
\\In close analogy to the computations in \cite{Calabrese:2004eu}, we first calculate the $n$th Renyi entropies $S^{(n)}_A=\frac{1}{1-n}\ln\mathrm{Tr} \rho^n_A$ where $\rho_A$ is the reduced density matrix of the state of the system where all degrees of freedom of the complement $B$ have been traced out. Evaluating this trace is equivalent to calculating a partition function $Z_n(A)$ on $n$ copies of the GCFT plane sewed together in an appropriate fashion. In a standard Euclidean CFT context this has an analytic meaning of an $n$-sheeted Riemann surface with a branch cut. From this one can the determine then entanglement entropy as a limit $S_E = - \lim_{n \rightarrow 1} \dfrac{\partial}{\partial n} \mathrm{Tr} \rho^n_A$. A mapping from the complex plane to this $n$-sheeted Riemann surface can be achieved via the conformal map
$z \rightarrow w= \left( \frac{z-u}{z-w}\right)^{1/n}$, where $u,v$ are the end points of $A$.\\
In the nonrelativistic setting of a GCFT, one has to construct an analogous $n$-sheeted surface $\Sigma_n$ as well. We first note that a map between $\Sigma_n$ and the GCFT plane can be established by the coordinate transformations,
\bes
\bea
 t &=& \left(\dfrac{t^{\prime}-t_1}{t^{\prime}-t_2} \right)^{1/n}, \\ 
x &=& \frac{1}{n}\left(\dfrac{t^{\prime}-t_1}{t^{\prime}-t_2} \right)^{1/n} \left( \dfrac{x^{\prime}-x_1}{t^{\prime}-t_1} - \dfrac{x^{\prime}-x_2}{t^{\prime}-t_2}  \right). 
\eea
\ees
The form of the transformed energy-momentum tensor components on $\Sigma_n$ can then be determined using \eqref{em_trans}. The vacuum expectation values of the energy-momentum tensors thus take the following form:
\begin{subequations} \label{Texp}
	\begin{align} 
\langle T_{(1)}(t^{\prime}, x^{\prime})\rangle _{\Sigma_n} =& \left(1-\frac{1}{n^2}\right)\left(\frac{t_{12}}{t_{{\prime}1} t_{{\prime}2}}\right)^2\times\nonumber\\
&\times\left[ \frac{c_L}{24} - \frac{c_M}{12} \left(\frac{x_{12}}{t_{12}}-\frac{x_{{\prime}1}}{t_{{\prime}1}}-\frac{x_{{\prime}2}}{t_{{\prime}2}}\right)\right],  \\
\langle T_{(2)}(t^{\prime}, x^{\prime})\rangle _{\Sigma_n} =& \left(1-\frac{1}{n^2}\right)\left(\frac{t_{12}}{t_{{\prime}1} t_{{\prime}2}}\right)^2 \frac{c_M}{24},
	\end{align}
\end{subequations} 
where $t_{\prime 1} = t^{\prime}- t_1, t_{12}= t_1-t_2,\ldots$. Here we have used the fact that the planar energy-momentum tensor has vanishing vacuum expectation value due to its symmetries. One can now compare the explicitly evaluated Ward identities on the GCFT plane \eqref{ward_exp} and the above vacuum expectation values \eqref{Texp} on $\Sigma_n$. This comparison yields
\bea \label{final}
\langle T_{(i)}(t^{\prime}, x^{\prime})\rangle _{\Sigma_n} = \dfrac{\langle T_{(i)}(x^{\prime},t^{\prime}) \Phi_n^{(1)} \Phi_{-n}^{(2)}\rangle_{\mathrm{Pl}}}{\langle\Phi_n^{(1)} \Phi_{-n}^{(2)}\rangle_{\mathrm{Pl}} },
\eea
for $i=1,2$, provided we identify the weights of the twist primaries $\Phi_n$ as \mbox{$h_L = \frac{c_L}{24} \left(1-\frac{1}{n^2}\right)$} and \mbox{$h_M = \frac{c_M}{24} \left(1-\frac{1}{n^2}\right)$}.\\
As mentioned above evaluating the quantity of interest, i.e., $\mathrm{Tr} \rho^n_A$, is the same as doing a path integral over $ \Sigma_n$ (up to a normalization). The left-hand side of the identity \eqref{final} is just a $T_{(i)}$ insertion in that functional integral. This insertion is equivalent to a conformal transformation defined by the Ward identities \eqref{ward2} and \eqref{ward1}. The functional integral on $\Sigma_n$ is therefore proportional to $n$ products of the GCFT plane two-point correlators of the twist fields evaluated at the end points of $A$: $\left(\langle \Phi_n (x_1, t_1) \Phi_{-n} (x_2, t_2)\rangle \right)^n$. This allows one to infer
	\begin{align}
		\mathrm{Tr} \rho^n_A=&k_n\left(\langle \Phi_n (x_1, t_1) \Phi_{-n} (x_2, t_2)\rangle \right)^n_\mathbb{C}\nonumber\\
		=&k_nt_{12}^{-\frac{c_L}{12}(n-\frac{1}{n})}\exp\left[\frac{c_M}{12}\left(n-\frac{1}{n}\right)\frac{x_{12}}{t_{12}}\right],
	\end{align}
where the $k_n$ are some normalization constants which we choose in a convenient way.\\
The final step is to determine the entanglement entropy $\mathrm{Tr} (\rho_A \, \ln \rho_A)$ for the segment $A$ as a limit of Renyi entropies. Taking the limit $n\rightarrow1$ of ${\frac{\partial}{\partial n}\mathrm{Tr}\rho^n_A}$ one obtains
\begin{equation} \label{final_2}
S_E =- \lim_{n \rightarrow 1} \dfrac{\partial}{\partial n} \mathrm{Tr} \rho^n_A = \frac{c_L}{6} \ln \left(\frac{t_{12}}{a}\right) + \frac{c_M}{6} \left( \frac{x_{12}}{t_{12}}\right),
\end{equation}
where $a$ is very small and is interpreted as a small scale cutoff or lattice spacing in the underlying GCFT. We will explain how to interpret this result and its physical content in light of holography in the next section.

\subsection{The BMS/GCA Correspondence and Entanglement Entropy}\label{sec:EEinGCFTCalc}

In Sec. \ref{GCFT} we laid the foundation for calculations in a 2D quantum field theory which is invariant under Galilean transformations in addition to having scale invariance. We now elaborate on how the results obtained previously are related to flat space holography in $2+1$ dimensions.\\
The asymptotic symmetries of an asymptotically flat spacetime at null infinity in $2+1$ bulk dimensions are known to be generated by an infinite but countable set of vector fields.
The infinite dimensional Lie algebra generated by these vector fields or rather their canonical charge algebra is known as the Bondi--van der Burg--Metzner--Sachs ($\mathfrak{bms}_3$) algebra, which is isomorphic to the GCA. However, the fact that an asymptotically flat spacetime can be viewed as a zero cosmological constant limit of an AdS spacetime flat holography should also have an interpretation as a limit of the AdS/CFT correspondence.\\
The asymptotic symmetries of $2+1$ dimensional pure Einstein asymptotically AdS spacetimes which consist of two copies of the Virasoro algebra with equal central charges $c=\bar{c} = \frac{3\ell}{2G}$ where $\ell$ is the AdS radius and $G$ is Newton's constant in $2+1$ dimensions. An asymptotically flat space-time can then be obtained from an AdS one by taking $\ell \rightarrow \infty$. This limit can be taken at various stages of the asymptotic symmetry analysis and it induces an \.In\"on\"u--Wigner contraction in the pair of Virasoro algebras formed by the asymptotic charges. To be more precise, the $\mathfrak{bms}_3$ generators are related to the AdS ones via
\bea \label{bms_contrac}
L_n = \lim_{\e \rightarrow 0} \left(\L_n - \bL_{-n} \right), ~ M_n = \lim_{\e \rightarrow 0}\e \left( \L_n + \bL_{-n}\right),~
\eea
with ${\e = \frac{G}{\ell}}$. 
Although the contraction is different in comparison to \eqref{nr}, the BMS generators $L_n, M_n$ form an algebra which is isomorphic to \eqref{gcaex}. For pure Einstein gravity one finds that $c_L =0, c_M = 3$. This is consistent with the definition corresponding to the $\ell \rightarrow \infty$ limit.\\
We now focus on the two-dimensional field theory perspective of this new contraction. Consider again the vector fields \eqref{witt_vf} of 2D conformal transformations. In the Lorentzian signature, i.e. with $z= t+x, \bar{z} = t-x$, the rescaling of the coordinates $x$ and $t$ which is consistent with the new contraction \eqref{bms_contrac} is given by \eqref{scaling2}. This corresponds to an ultrarelativistic contraction of the Virasoro symmetries.\\
The crucial point of this discussion is that the nonrelativistic (Galilean) and the ultrarelativistic limits of the 2D conformal algebra are algebraically isomorphic. This is a property of two dimensions because 1-space dimension and 1-time direction can always interchange their roles since the parent theory has special relativistic invariance which does not distinguish between coordinates.\\
Taking this BMS/GCA correspondence into account, all the field theory results derived earlier in this paper in the context of GCFT can be used for calculations for ultrarelativistic field theories invariant under $\mathfrak{bms}_3$ symmetries at null infinity of asymptotically flat spacetimes by only exchanging the role of time and space.\\
Using this argument one can immediately determine entanglement entropy in a planar field theory having ultrarelativistic conformal/$\mathfrak{bms}_3$ symmetry. Once again we take a rectilinear segment $A$ with end points $(x_1, t_1)$ and $(x_2, t_2)$ as our entangling region. The entanglement entropy of that region can readily be written using \eqref{final_2} and exchanging time and space
\bea \label{bms_entropy}
S_E  = \frac{c_L}{6} \ln \left(\frac{x_{12}}{a}\right) + \frac{c_M}{6} \left( \frac{t_{12}}{x_{12}}\right).
\eea
This interval can again be interpreted as a boosted version of a purely spatial (or equal time) interval. For $t_{12}=0$ \eqref{bms_entropy} reduces to
\bea \label{equal_time}
 S_E  = \frac{c_L}{6} \ln \left(\frac{x_{12}}{a}\right).
\eea
The entanglement entropy calculated above corresponds to a two-dimensional system of infinite spatial extent at zero temperature. It is also of interest to see what happens with the entanglement entropy when dealing with a system at finite temperature and/or finite spatial extent. Thus, in the following paragraphs we generalize the result \eqref{bms_entropy} to a system at finite temperature $T= \beta^{-1}$.\\
This generalization can be achieved by using geometric properties of the two-dimensional field theory in question. To elucidate this a bit more we note that one can map the two-dimensional GCFT on the plane to a cylinder by
\begin{equation}
	x= e^{2 \pi \xi /\beta},\quad t= \frac{2 \pi \tau}{\beta}e^{2 \pi \xi /\beta},
\end{equation}
where $\xi$ and $\tau$ denote the coordinates on the cylinder. This means that effectively one dimension gets compactified in the construction of $\Sigma_n$. As shown in \cite{Bagchi:2013qva} this induces a transformation of the GCFT primaries as
\bea
\tilde{\Phi} (\xi, \tau) = e^{\frac{2 \pi}{\beta}( \xi h_L +  \tau h_M)} \Phi (x(\xi, \tau), t(\xi, \tau)).
\eea
The two-point function evaluated in this geometry is then given by
\bea
\langle \tilde{\Phi} (\xi_1, \tau_1)  \tilde{\Phi} (\xi_2, \tau_2)\rangle =& \left [ 2 \sinh\left(\frac{\pi\xi_{12}}{\beta}\right)\right]^{-2h_L}\times\nonumber\\
&\times e^{-h_M\frac{\pi\tau_{12}}{\beta}. \coth\left(\frac{\pi \xi_{12}}{\beta}\right)}
\eea
The following steps which are necessary for calculating entanglement entropy in a thermal state for a subsystem with end points $\xi_1, \tau_1$ and $\xi_2, \tau_2$ are the same as in the zero-temperature case. Thus one obtains for the entanglement entropy for a system at finite temperature the following expression:
\bea \label{ee_temp}
S_E = \frac{c_L}{6} \ln \left[\frac{\beta}{\pi a} \sinh \left( \frac{\pi \xi_{12}}{\beta}\right) \right] + \frac{\pi c_M}{6 \beta} \tau_{12} \coth \left( \frac{\pi \xi_{12}}{\beta}\right)~
\eea
This result can be further analyzed  by considering its low-temperature and high-temperature limits.\\
At leading order the expansion of the right-hand side of \eqref{ee_temp} in $\beta^{-1}$ yields again--as expected--the zero-temperature answer \eqref{bms_entropy} with the identification of $\tau_{12} \sim t_{12}/a$ and $\xi_{12} \sim x_{12}/a$. In the high temperature limit, on the other hand, i.e., for $ \xi_{12} \gg \beta$ one obtains
\bea \label{ee_hightemp}
S_E = \frac{\pi}{6 \beta} \left( c_L \xi_{12} + c_M \tau_{12}\right) + \frac{c_L}{6} \ln \beta + \mathcal{O}(\beta).
\eea
A very similar analysis works when considering the spatial extent of the system to be of finite length $L$ in the ground state. The only difference in comparison to the analysis before lies in the direction of the compactification to the cylinder along the spatial cycle of length $L \sim \beta$ which is perpendicular to the previous case. The entanglement entropy for that system then turns out to be
\bea \label{finite_system}
S_E = \frac{c_L}{6} \ln \left[\frac{L}{\pi a} \sin \left(\frac{\pi \xi_{12}}{L}\right) \right] + \frac{\pi c_M}{6 L} \tau_{12} \cot \left( \frac{\pi \xi_{12}}{L}\right).~~
\eea

\section{Thermal Entropy in GCFTs}\label{sec:ThermalEntropy}

In this section we briefly review how to derive the high-temperature density of states and the corresponding entropy for ordinary 2D GCFTs (for more details see, e.g., Refs. \cite{Bagchi:2012xr,Barnich:2012xq,Bagchi:2013qva}) in order to make contact with the holographic results for the thermal entropy of FSCs found in Sec. \ref{sec:ThermalEntropyFSC}.\\
The partition function for a 2D GCFT on a torus is given by
	\begin{align}\label{eq:PartFunct}
		Z^0_{GCFT}(\eta,\rho)=&\textnormal{Tr}\left(e^{2\pi i\eta\left(L_0-\frac{c_L}{24}\right)}e^{2\pi i\rho\left(M_0-\frac{c_M}{24}\right)}\right)\nonumber\\
		=&e^{\frac{\pi i}{12}\left(\eta c_L+\rho c_M\right)}Z_{GCFT}(\eta,\rho),
	\end{align}
where $\eta$ and $\rho$ are the Galilean conformal equivalents of the modular parameters of a CFT defined on a torus. In the same spirit as in a relativistic CFT, one demands that \eqref{eq:PartFunct} is invariant under the Galilean conformal equivalent of $S$ modular transformations given by
	\begin{equation}\label{eq:STransform}
		\left(\eta,\rho\right)\rightarrow\left(-\frac{1}{\eta},\frac{\rho}{\eta^2}\right),
	\end{equation}
i.e.,	
	\begin{equation}
		Z^0_{GCFT}(\eta,\rho)=Z^0_{GCFT}(-\frac{1}{\eta},\frac{\rho}{\eta^2}).
	\end{equation}
This is tantamount to requiring
	\begin{equation}
		Z_{GCFT}(\eta,\rho)=e^{2\pi i\left(\tilde{f}(\eta,\rho)+h_L\eta+h_M\rho\right)}Z_{GCFT}(-\frac{1}{\eta},\frac{\rho}{\eta^2}),
	\end{equation}
where
	\begin{equation}
		\tilde{f}\left(\eta,\rho\right)=\frac{c_L\eta}{24}+\frac{c_M\rho}{24}+\frac{c_L}{24\eta}-\frac{c_M\rho}{24\eta^2}-h_L\eta-h_M\rho.
	\end{equation}
In order proceed one first has to rewrite the density of states $d\left(h_L,h_M\right)$ in terms of the GCA partition function. This can be done by using	
	\begin{align}
		Z_{GCFT}(\eta,\rho)=&\textnormal{Tr}\left(e^{2\pi i\eta L_0}e^{2\pi i\rho M_0}\right)\nonumber\\
		=&\sum d\left(h_L,h_M\right)e^{2\pi i\eta h_L}e^{2\pi i\rho h_M},
	\end{align}
and performing an inverse Laplace transformation
	\begin{equation}\label{eq:DensOfStates}
		d\left(h_L,h_M\right)=\int\extd\eta\extd\rho e^{2\pi i\tilde{f}\left(\eta,\rho\right)}Z_{GCFT}(-\frac{1}{\eta},\frac{\rho}{\eta^2}).
	\end{equation}
In the limit of large central charges the density of states \eqref{eq:DensOfStates} can by approximated by the value of the integrand, when the exponential factor is extremal. Using this approximation the density of states is given by
	\begin{equation}
		d\left(h_L,h_M\right)\sim e^{\pi\,\sqrt{\frac{c_Mh_M}{6}}\,\left(\frac{h_L}{h_M}+\frac{c_L}{c_M}\right)}.
	\end{equation}
The corresponding entropy is then simply given by the logarithm of the density of states,
	\begin{equation}\label{eq:PartGCAEntropy}
		S=\ln\left(d\left(h_L,h_M\right)\right)=\pi\,\sqrt{\frac{c_Mh_M}{6}}\,\Bigg(\frac{h_L}{h_M}+\frac{c_L}{c_M}\Bigg).
	\end{equation}

\section{Flat Space Entanglement Entropy}\label{sec:HFSEE}

In this part of our paper we describe how to determine entanglement entropy holographically by employing a modified version of the Wilson line approach first presented in \cite{Ammon:2013hba,Castro:2014tta}.

\subsection{Wilson Lines in AdS$_3$ Representing a \emph{Massive} and \emph{Spinning} Particle}
In order to find a suitable proposal for holographic entanglement entropy in flat space, we modify the proposal made in \cite{Ammon:2013hba,Castro:2014tta}\footnote{Another proposal for holographic entanglement entropy using Wilson lines was also made in \cite{deBoer:2013vca}. In \cite{Castro:2014mza} it was shown that this proposal is equivalent to the proposal in \cite{Ammon:2013hba,Castro:2014tta}.} for AdS$_3$. In the following section we review briefly the main concepts underlying this proposal.\\
Einstein gravity in 2+1 dimensions can conveniently be reformulated as a Chern-Simons gauge theory \cite{Achucarro:1987vz,Witten:1988hc} with the action
	\begin{equation}\label{eq:ChernSimonsAction}
		I_{CS}[\mathcal{A}]=\frac{k}{4\pi}\int\left<\mathcal{A}\wedge\extd\mathcal{A}+\frac{2}{3}\mathcal{A}\wedge\mathcal{A}\wedge\mathcal{A}\right>,
	\end{equation}
where $k$ is the Chern-Simons level, $\mathcal{A}$ takes values in some gauge algebra $\mathfrak{g}$ and $\left<\ldots\right>$ denotes a suitable invariant
bilinear form on the gauge algebra. If one describes AdS$_3$ in terms of this Chern-Simons formulation then the gauge field $\mathcal{A}$ takes values in $\mathfrak{so}(2,2)\sim\mathfrak{sl}(2,\mathbb{R})\oplus\mathfrak{sl}(2,\mathbb{R})$ and one can conveniently split the gauge field as follows
	\begin{equation}\label{eq:AdS3GaugeFieldSplit}
		\mathcal{A}=\left(
			\begin{array}{cc}
				 A_L & 0 \\
				 0 & A_R\\
			\end{array}\right),
	\end{equation}
with $A_L,A_R\in\mathfrak{sl}(2,\mathbb{R})$.\\
In \cite{Ammon:2013hba,Castro:2014tta} the authors argued that a Wilson line $W_\mathcal{R}(C)$, for an appropriate choice of representation $\mathcal{R}$, attached to the boundary of AdS$_3$ can be used to determine the entanglement of the region bounded by the end points of the Wilson line holographically as
	\begin{equation}\label{eq:AdSSEEWilsonLineRelation}
		S_{\textnormal{EE}}=-\log\left[W_\mathcal{R}(C)\right].
	\end{equation}
In the case that the central charges $c$ and $\bar{c}$ of the two copies of the Virasoro algebra at the boundary of AdS$_3$ are not equal i.e. $c\neq\bar{c}$ this Wilson line describes a massive and spinning particle\footnote{If $c=\bar{c}$ then the Wilson line describes a massive particle without spin.} probing the bulk geometry. Thus for a massive and spinning particle in AdS$_3$, where one can write $\mathcal{A}$ as in \eqref{eq:AdS3GaugeFieldSplit} one can also split the Wilson line accordingly as follows
	\begin{equation}
		W_\mathcal{R}(C)=W^L_\mathcal{R}(C)\times W^R_\mathcal{R}(C),
	\end{equation}
where
	\begin{align}\label{eq:AdSWilsonLineDefTrAndPathInt}
		W^{L}_\mathcal{R}(C)=&\textnormal{Tr}_\mathcal{R}\left(\mathcal{P}\exp\left(\int_CA_{L}\right)\right)\nonumber\\
		=&\int\mathcal{D}U_{L}\exp\left(-S_{L}(U_{L};A_{L})_C\right),
	\end{align}
and $A_{L}$ is the pullback of the connections along the curve $C$, i.e. $A=A_\mu\dot{x}^\mu$. The relevant expressions for ${W^{L}_\mathcal{R}}$ can be obtained by a simple exchange of the labels as ${L\leftrightarrow R}$.
The corresponding actions for left and right movers are given by
	\begin{subequations}\label{eq:SLSRAdS3}
		\begin{align}
			S_{L}(U_{L};A_{L})_C=&\int_C\extd s\left<P_LD_LU_LU_L^{-1}\right>+\lambda_L\left(\left<P_L^2\right>-c_2\right),\\	
			S_{R}(U_{R};A_{R})_C=&\int_C\extd s\left<P_RU_R^{-1}D_RU_R\right>+\lambda_R\left(\left<P_R^2\right>-\bar{c}_2\right),
		\end{align}
	\end{subequations}
where $c_2$ and $\bar{c_2}$ are the quadratic Casimirs of the two $\mathfrak{sl}(2,\mathbb{R})$ copies, $\left<\ldots\right>$ corresponds to the invariant bilinear form on each of the $\mathfrak{sl}(2,\mathbb{R})$ algebras, $U_{L}$ ($U_{R}$) describes the probe and takes values in the group manifold SL$(2,\mathbb{R})$ and $P_{L}$ ($P_{R}$) are the canonical momenta associated with $U_{L}$ ($U_{R}$) which take values in the Lie algebra $\mathfrak{sl}(2,\mathbb{R})$. The general strategy to determine the holographic entanglement entropy using these ingredients can be roughly summarized as follows:
	\begin{itemize}
		\item Determine the equations of motion (EOM) of \eqref{eq:SLSRAdS3}.
		\item Solve EOM with $A_{L}$ ($A_{R}$) set to zero (``nothingness trick'').
		\item Use a suitable (large) gauge transformation in order to generate a nontrivial solution of interest.
		\item Determine the path integral in \eqref{eq:AdSWilsonLineDefTrAndPathInt} using a saddle-point approximation.
		\item Use \eqref{eq:AdSSEEWilsonLineRelation} to determine the holographic entanglement entropy.
	\end{itemize}
\subsection{Constructing a Topological Probe for Flat Space}
Having recapitulated the basic ingredients of the holographic entanglement entropy proposal using Wilson lines in AdS$_3$ in the previous subsection, we now proceed in constructing a topological probe for flat space. For flat space in 2+1 dimensions one can again formulate the gravity theory in terms of a Chern-Simons action but the gauge field $\mathcal{A}$ now takes values in $\mathfrak{isl}(2,\mathbb{R})$ which is a semidirect sum $\mathfrak{sl}(2,\mathbb{R})\oplus_s\mathbb{R}^3$. The correct invariant bilinear form which has to be used in \eqref{eq:ChernSimonsAction} is given by \cite{Witten:1988hc,Afshar:2013vka}
	\begin{equation}\label{eq:ISL2InvBilinearForm}
		\left< L_n,L_m\right>=0,\,\left< L_n,M_m\right>=-\frac{1}{2}\eta_{nm},\,\left< M_n,M_m\right>=0,
	\end{equation}
where $\eta_{nm}=\textnormal{antidiag}(1,-\frac{1}{2},1)$ and $m,n=\pm1,0$. For some aspects of the following calculations in this part of the paper it will prove useful to find ways of writing \eqref{eq:ISL2InvBilinearForm} in terms of traces of a given matrix representation of $\mathfrak{isl}(2,\mathbb{R})$. This can be done, for example, by using the Grassmann approach described in detail in \cite{Krishnan:2013wta}. In this approach one interprets the AdS radius in a flat space limit as a Grassmann valued parameter $\epsilon$ with $\epsilon^2=0$. This can then be used to write down a very convenient matrix representation of $\mathfrak{isl}(2,\mathbb{R})$. See the Appendix for explicit expressions of the corresponding generators. In addition one can define new traces which have very advantageous properties for the purpose of our calculations. Another reason to use this approach and the matrix representation derived thereof is that one has a very direct and simple way of distinguishing quantities that derive from the $\mathfrak{sl}(2,\mathbb{R})$ (even) part of $\mathfrak{isl}(2,\mathbb{R})\sim\mathfrak{sl}(2,\mathbb{R})\oplus_s\mathbb{R}^3$ from the ones that derive from the $\mathbb{R}^3$ (odd). This is done by properly keeping track of powers of $\epsilon$. Terms which do not contain powers of $\epsilon$ are what we call even, and terms which are linear in $\epsilon$ are odd.\footnote{Since $\epsilon^2=0$ all powers of $\epsilon$ which are quadratic or higher vanish.}\\
In the previous section we mentioned that aside from a curve $C$ one also has to choose an appropriate representation $\mathcal{R}$. For AdS$_3$ the representation has to be chosen in such a way that the Wilson line corresponds to a massive and spinning particle moving in the AdS$_3$ bulk. As argued in \cite{Ammon:2013hba} one possible choice for this representation is an infinite dimensional highest-weight representation of $\textnormal{SL}(2,\mathbb{R})\otimes \textnormal{SL}(2,\mathbb{R})$ characterized by the conformal weights $(h,\bar{h})$. In close analogy to this we claim that for flat space the correct choice of representations is an infinite dimensional highest-weight representation of ISL$(2,\mathbb{R})$ characterized by the Galilean conformal weights $(h_L,h_M)$. We implement these representations in the exact same way as described in \cite{Ammon:2013hba} i.e. we construct an auxiliary quantum mechanical system defined on the Wilson line whose Hilbert space will be exactly the representation $\mathcal{R}$ we want.\\
The calculations performed in Sec. \ref{sec:EEinGCFTCalc} show that the entanglement entropy for GCFTs splits into two different parts which are proportional to the central charges $c_L$ and $c_M$. Thus, it seems natural that, similar to the AdS$_3$ case, we mimic that behavior by splitting the action $S(U;\mathcal{A})_C$ appearing in the path integral 
	\begin{align}
		W_\mathcal{R}(C)=&\textnormal{Tr}_\mathcal{R}\left(\mathcal{P}\exp\left(\int_C\mathcal{A}\right)\right)\nonumber\\
		=&\int\mathcal{D}U\exp\left(-S(U;\mathcal{A})_C\right),
	\end{align}
which determines the Wilson line and is used to construct the auxiliary quantum system, into even and odd parts labeled by $L$ and $M$, respectively, and we also fix the norm of the canonical momenta in a similar manner. This in turn also means that the holographic entanglement entropy written in terms of Wilson lines should be given by
	\begin{equation}\label{eq:SEEFlatSplitDefinition}
		S_{\textnormal{E}}=-\log\left[W_\mathcal{R}^L(C)\right]-\log\left[W_\mathcal{R}^M(C)\right].
	\end{equation}
In order to proceed with this split we first assume that one can write the topological probe $U$ and $S(U;\mathcal{A})_C$ as
	\bea\label{eq:SplitAssumtion}
		U\in \textnormal{ISL}(2,\mathbb{R}),\quad\textnormal{and}\quad U_{L+M}=U_LU_M,\non \\
		S(U;\mathcal{A})_C=S_L(U_L;A_L)_C+S_M(U_M;A_M)_C,
	\eea
with
	\begin{subequations}\label{eq:SLSMFlat}
		\begin{align}
			S_L=&\int_C\extd s\left<P_LD_LU_LU_L^{-1}\right>_L+\lambda_L\left(\left<P_L^2\right>_L-c_2\right),\label{eq:SLFlat}\\
			S_M=&\int_C\extd s\left<P_MD_MU_MU_M^{-1}\right>_M+\lambda_M\left(\left<P_M^2\right>_M-\bar{c}_2\right),\label{eq:SMFlat}
		\end{align}
	\end{subequations}
where $s\in[0,1]$ parametrizes the curve $C$ and
	\begin{equation}
		D_{L}=\partial_s+A_{L},\quad D_{M}=\partial_s+A_{M}
	\end{equation}
 $P_{L}$ $(P_M)$ is the canonical momentum conjugate to $U_{L}$ $(U_M)$, and $\lambda_{L}$ $(\lambda_M)$ is a Lagrange multiplier that constrains the norm of $P_{L}$ $(P_M)$ to $c_2$ or $\bar{c}_2$ respectively.\footnote{For flat space $c_2$ is the value of the quadratic casimir operator of the $\mathfrak{sl}(2,\mathbb{R})$ part of $\mathfrak{isl}(2,\mathbb{R})$ i.e. $c_2=2L_0^2-\left(L_{-1}L_1+L_{1}L_{-1}\right)$ while $\bar{c}_2$ is one of the quadratic casimirs of the \emph{full} $\mathfrak{isl}(2,\mathbb{R})$ algebra (the other one would be the helicity) i.e. $\bar{c}_2=2M_0^2-\left(M_{-1}M_1+M_{1}M_{-1}\right)$ which label the representation $\mathcal{R}$ via the highest weights $h_L$ and $h_M$.} The invariant bilinear forms $\left<\ldots\right>_{L}$ ($\left<\ldots\right>_{M}$) can also be written in terms of Lie algebra metrics $\omega_{ab}$ and $\bar{\omega}_{ab}$, where $\omega_{ab}$ is restricted to the even and $\bar{\omega}_{ab}$ to odd generators as follows:
 	\begin{subequations}
 		\begin{align}
 			\left<P^2\right>_{L}=&P_aP_b\omega^{ab}=2P_0^2-\left(P_{-1}P_1+P_{1}P_{-1}\right),\\
 			\left<\bar{P}^2\right>_{M}=&\bar{P}_a\bar{P}_b\bar{\omega}^{ab}=2\bar{P}_0^2-\left(\bar{P}_{-1}\bar{P}_1+\bar{P}_{1}\bar{P}_{-1}\right),
 		\end{align}
	\end{subequations} 
for $P=P_aL^a$ and $\bar{P}=\bar{P}_aM^a$. The even metric $\omega_{ab}$ can be determined using the ordinary trace and the matrix representation found in the Appendix (with a factor of 1/2), i.e., $\omega_{ab}=\frac{1}{2}\textnormal{Tr}\left(L_aL_b\right)$. In order to determine $\bar{\omega}_{ab}$ a ``twisted'' trace which is defined as
	\begin{equation}\label{eq:TwistedTraceDefinition}
		\bar{\textnormal{Tr}}\left(\prod_{i=1}^k\mathcal{G}_i\right)=\frac{1}{2}\textnormal{Tr}\left(\prod_{i=1}^k\frac{\extd}{\extd \epsilon}\mathcal{G}_i\gamma^\star_{(2)}\right),
	\end{equation}
with
	\begin{equation}\label{eq:GammaStarI}
	\gamma^\star_{(a)}=\left(
		\begin{array}{cc}
			\unity_{a\times a} & 0 \\
			0 & -\unity_{a\times a}\\
		\end{array}\right),
	\end{equation}	
where $\mathcal{G}_i\in\mathfrak{isl}(2,\mathbb{R})$ can be used in a similar way i.e. $\bar{\omega}_{ab}=\bar{\textnormal{Tr}}\left(M_aM_b\right)$.\\ 
The EOMs for the even part of \eqref{eq:SLSMFlat} are given by
	\begin{equation}\label{eq:SLSMFlatEOM}
		D_{L}U_{L}U_{L}^{-1}+2\lambda_{L}P_{L}=0,\quad \frac{\extd}{\extd s}P_{L}=0,
	\end{equation}
in addition to the constraints $\left< P_L^2\right>_L=c_2$. The EOMs for the odd part are the same as in \eqref{eq:SLSMFlatEOM} upon replacing ${L\leftrightarrow M}$ in addition to the constraint $\left< P_M^2\right>_M=\bar{c}_2$. In order to solve these EOMs one can use the same ``nothingness trick'' as in the AdS$_3$ case by finding a solution of these equations first for $A_{L}=0$ ${(A_{M}=0)}$ and then generating a nontrivial solution by using a (large) gauge transformation. For $A_{L}=0$ ${(A_{M}=0)}$ solutions of \eqref{eq:SLSMFlatEOM} are given by
	\begin{align}\label{eq:U0Solution}
		U^{(0)}_{L}=&u_{L}^{(0)}\exp\left(-2\alpha_{L}(s)P_{L}^{(0)}\right),\nonumber\\
		\frac{d\alpha_{L}(s)}{ds}=&\lambda_{L}(s),\quad (L\leftrightarrow M),
	\end{align}
where $u_{L}^{(0)}$ $(u_{M}^{(0)})$ are constant group elements chosen in such a way that \eqref{eq:SplitAssumtion} is satisfied. Looking at \eqref{eq:U0Solution} and the assumption  \eqref{eq:SplitAssumtion}, one also finds that, $\left[P_{L}^{(0)},P_{M}^{(0)}\right]=0$ has to be satisfied. Using this one obtains the following on-shell actions:
	\begin{equation}
		S_{L}^{\textnormal{on-shell}}=-2\Delta\alpha_{L}c_2,\quad S_{M}^{\textnormal{on-shell}}=-2\Delta\alpha_{M}\bar{c}_2,
	\end{equation}
where $\Delta\alpha_{L}=\alpha_{L}(1)-\alpha_{L}(0)$ and equivalently for $\Delta\alpha_{M}$. By using a saddle-point approximation for the path integral 
	\begin{equation}
		\int\mathcal{D}Ue^{-S(U;\mathcal{A})_C}\sim e^{-S_{\textnormal{on-shell}}(U;\mathcal{A})_C},
	\end{equation}
one can write \eqref{eq:SEEFlatSplitDefinition} as
	\begin{equation}
		S_{\textnormal{EE}}=-2\Delta\alpha_{L}c_2-2\Delta\alpha_{M}\bar{c}_2.
	\end{equation}
Thus the calculation of holographic entanglement entropy using Wilson lines reduces to calculating $\Delta\alpha_{L}$ and $\Delta\alpha_{M}$ for the relevant theories in question.

\subsection{Calculating Holographic Entanglement Entropy for Flat Space}

\begin{figure}[h]
		\centering
		\includegraphics[width=0.5\textwidth]{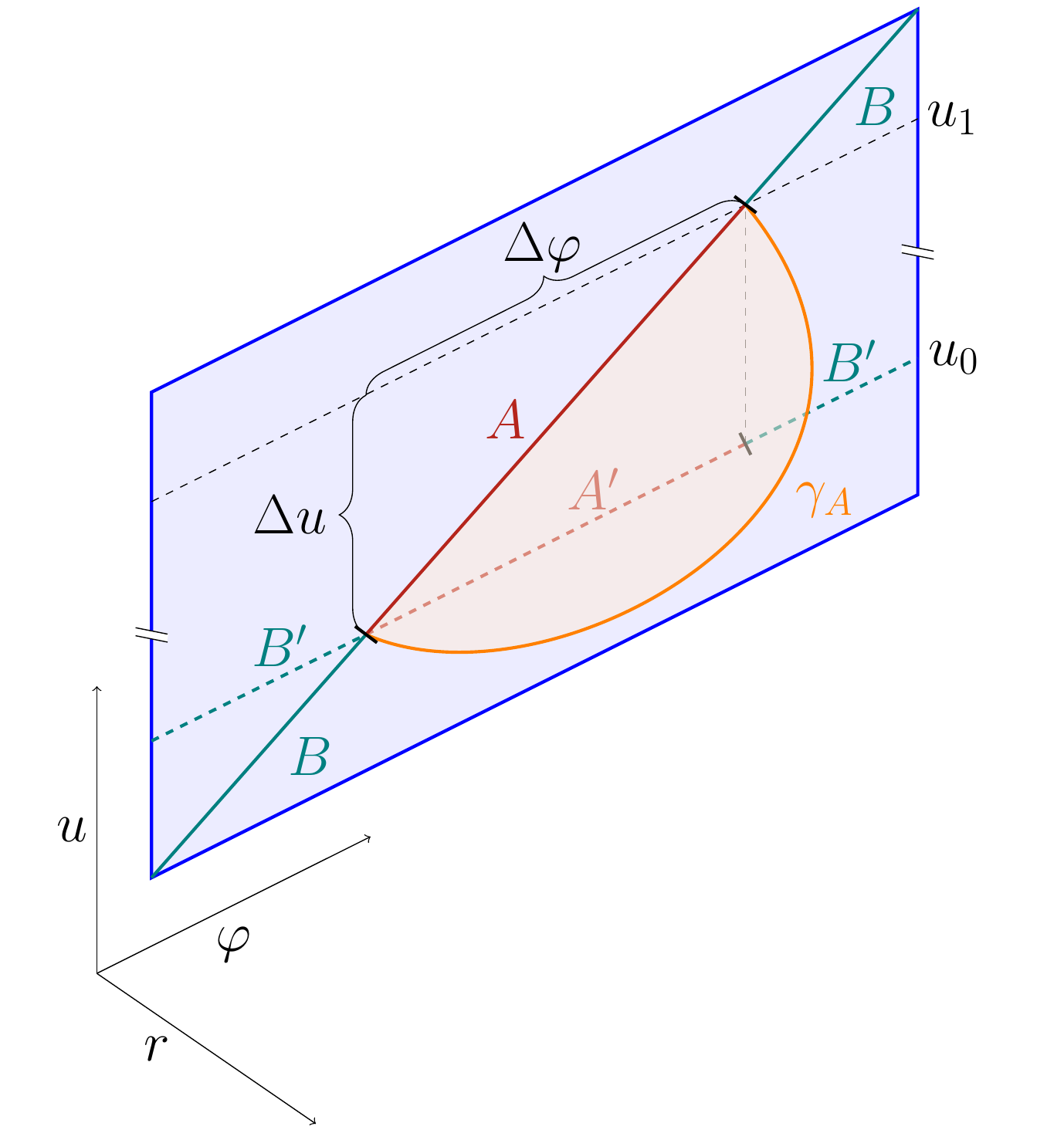}
		\caption{Boosted ({\color{BrickRed}$A$}, {\color{teal}$B$}) and equal time ({\color{BrickRed}$A'$}, {\color{teal}$B'$}) entangled intervals and the correpsonding Wilson line ({\color{orange}$\gamma_A$}) used to determine holographic entanglement entropy in flat space.}
		\label{fig:HEEPicture}
	\end{figure}

Having constructed a suitable topological probe for flat space in the previous subsection, we now calculate the entanglement entropy for various different flat spacetimes holographically.\\
The most general solutions of 2+1-dimensional flat space Einstein gravity with coordinates\footnote{We assume that the topology of our spacetime manifold is given by a solid cylinder, where $r$ is a radial coordinate with a boundary located at $r=\infty$, $u$ is the retarded time and $\varphi\sim\varphi+2\pi$ is an angular spacelike coordinate.} $r,u,\varphi$ is given by the line element \cite{Barnich:2012aw}
	\begin{equation}
		\extd s^2=\mathcal{M}(\varphi) \extd u^2-2\extd u\extd r+2\mathcal{N}(u,\varphi)\extd u\extd \varphi+r^2\extd\varphi^2,
	\end{equation}
where $\mathcal{M}$ and $\mathcal{N}$ have to satisfy
	\begin{equation}
		2\partial_u\mathcal{N}=\partial_\varphi\mathcal{M}.
	\end{equation}
For $\mathcal{M}=\mathcal{N}=0$ this solution is known as the null orbifold \cite{Horowitz:1990ap,FigueroaO'Farrill:2001nx,Liu:2002kb,Simon:2002ma}. For $\mathcal{M}=-1,\,\mathcal{N}=0$ the solution is (global) flat space, and the generic case $\mathcal{M}\geq0$ and $\mathcal{N}\neq0$, where $\mathcal{M}$ and $\mathcal{N}$ are constant, corresponds to flat space cosmological solutions \cite{Cornalba:2002fi,Cornalba:2003kd}.\\ 
As in the AdS$_3$ case it is convenient to formulate flat space gravity in terms of a Chern-Simons action and the corresponding gauge connection $\mathcal{A}$. In terms of this gauge connection the various solutions mentioned in the preceding paragraph can be written as \cite{Barnich:2014cwa,Gary:2014ppa}
	\begin{equation}
		\mathcal{A}=b^{-1}(\extd+a)b\quad\textnormal{with}\quad b=e^{\frac{r}{2}M_{-1}},
	\end{equation}
with
	\begin{align}\label{eq:Spin2ChernSimonsConnection}
		a=&\left(M_1-\frac{\mathcal{M}}{4}M_{-1}\right)\extd u\nonumber\\
		&+\left(L_1-\frac{\mathcal{M}}{4}L_{-1}-\frac{\mathcal{N}}{2}M_{-1}\right)\extd\varphi.
	\end{align}
As in the previous section we split this connection into an even and odd part, respectively, as ${\mathcal{A}=A_L+A_M}$ with 
	\begin{subequations}\label{eq:ALAMFSC}
		\begin{align}
			A_L=&\left(L_1-\frac{\mathcal{M}}{4}L_{-1}\right)\extd\varphi,\\
			A_M=&\frac{1}{2}M_{-1} \extd r+\left(M_1-\frac{\mathcal{M}}{4}M_{-1}\right)\extd u\nonumber\\
			&+\left(r M_0-\frac{\mathcal{N}}{2}M_{-1}\right)\extd\varphi.
		\end{align}
	\end{subequations}
One can now use the connections $A_L$ and $A_M$ and perform a large gauge transformation on the trivial solution \eqref{eq:U0Solution} in order to obtain a solution for \eqref{eq:SLSMFlatEOM} with $A_L$ and $A_M$ given by \eqref{eq:ALAMFSC}. This gauge transformation can be compactly written as
	\begin{equation}
		A_L+A_M=A\extd A^{-1}\quad\textnormal{with}\quad A=b^{-1}e^{-\int a_i\extd x^i}.
	\end{equation}
The topological probe $U(s)$ transforms under this gauge transformation as
	\begin{equation}\label{eq:UGaugedSolution}
		U(s)=\left(U_LU_M\right)(s)=A(s)U^{(0)}_{L}U^{(0)}_{M}A^{-1}(s).
	\end{equation}
Up until this point of the calculation it was not necessary to specify the exact points at which the Wilson line is attached. However, since we have to fix boundary conditions for our probe at some point during our calculations, we now specify where exactly the Wilson line is attached and to which entangling interval it is bounding (see Fig. \ref{fig:HEEPicture}). 
\\First we introduce a radial cutoff $r_0$ which is placed very close to the boundary $r=\infty$ in order to regulate infinities when approaching the boundary. The Wilson line will then be attached at the hypersurface with $r=r_0$ at the points $x^\mu_i=(r_0,u_i,\varphi_i)$ and $x^\mu_f=(r_0,u_f,\varphi_f)$. Denoting
	\begin{align}
		U(0)&=U_i,&& U(1)=U_f,\nonumber\\
		A(0)&=A\big|_{x=x_i}=A_i,&&  A(1)=A\big|_{x=x_f}=A_f,\non \\
		\alpha_{L}(0)&=\alpha_{L}^i,&&\alpha_{L}(1)=\alpha_{L}^f,\non\\
		\alpha_{M}(0)&=\alpha_{M}^i,&&\alpha_{M}(1)=\alpha_{M}^f,
	\end{align}
one can use \eqref{eq:UGaugedSolution} to write	
	\begin{subequations}
		\begin{align}
			U_{i}=A_{i}u_{L}^{(0)}u_{M}^{(0)}\exp\left(-2\alpha_{L}^{i}P_{L}^{(0)}-2\alpha_{M}^{i}P_{M}^{(0)}\right)A^{-1}_{i},\\
			U_{f}=A_{f}u_{L}^{(0)}u_{M}^{(0)}\exp\left(-2\alpha_{L}^{f}P_{L}^{(0)}-2\alpha_{M}^{f}P_{M}^{(0)}\right)A^{-1}_{f}.
		\end{align}
	\end{subequations}
Next solving for $u_{L}^{(0)}u_{M}^{(0)}$ in one of the two equations and replacing the expression in the remaining equation one obtains 
	\begin{equation}\label{eq:TheMightyEquation}
		e^{-2\Delta\alpha_{L}P_{L}^{(0)}-2\Delta\alpha_{M}P_{M}^{(0)}}=A_i^{-1}U_i^{-1}A_iA_f^{-1}U_fA_f=\Omega.
	\end{equation}
Having obtained this equation one can now almost determine $\Delta\alpha_{L}$ and $\Delta\alpha_{M}$. The only thing left to do is to choose appropriate boundary conditions for the topological probe at the initial and final points of the Wilson line. As in the AdS$_3$ case it is, as of yet, not known how unique such a choice of boundary conditions actually is, i.e., if there is only one set of boundary conditions that yields the correct entanglement entropy or if there is a whole family thereof. In the pure AdS$_3$ case for spin-2 one can employ boundary conditions, for example, in such a way that the curve the Wilson line is describing is actually a geodesic \cite{Ammon:2013hba}, in accordance with the Ryu-Takayanagi proposal. For other cases like the ones described in \cite{Castro:2014tta} where one has to deal with gravitational anomalies which render the theory non-Lorentz invariant the guiding principle is not so clear. In the case at hand we try to choose our boundary conditions in such a way that they are as simple as possible and analogous to the ones for theories with gravitational anomalies. The reason for this is that looking at flat space as a limit from AdS theories with gravitational anomalies can be seen as the ``parent'' theories for GCFTs with $c_M\neq0$. Following this reasoning we propose the following boundary conditions for the topological probe $U$ at the initial and final points	
	\begin{equation}
		U^{-1}_i=e^{\frac{r}{2}L_{-1}}b,\quad U_f=e^{-\frac{r}{2}L_{-1}}b.
	\end{equation}
After fixing the boundary conditions one can solve \eqref{eq:TheMightyEquation} for $\Delta\alpha_{L}$ and  $\Delta\alpha_{M}$.
In order to proceed, it makes sense to first take a closer look at \eqref{eq:TheMightyEquation} and use the fact that $\mathfrak{isl}(2,\mathbb{R})$ has a nilpotent subalgebra. Since ${\left[M_n,M_n\right]=0}$ and we assumed $\left[P_{L}^{(0)},P_{M}^{(0)}\right]=0$, Eq. \eqref{eq:TheMightyEquation} simplifies to
	\begin{equation}
		e^{-2\Delta\alpha_{L}P_{L}^{(0)}}\left(\unity-2\Delta\alpha_{M}P_{M}^{(0)}\right)=\Omega.
	\end{equation}	
At this point the way the $\mathfrak{isl}(2,\mathbb{R})$ matrix representation used is constructed and categorized in even and odd parts is again very convenient as one can schematically write the left-hand side of this equation as
	\begin{equation}\label{eq:EWSplit}
		\left(
			\begin{array}{cc}
			  e^{\gamma_L}&0\\
			0& e^{-\gamma_L}\\
			\end{array}\right)\otimes\unity_{2\times2}+\epsilon\gamma_M
			\left(
			\begin{array}{cc}
			 e^{\gamma_L}&0\\
			0& -e^{\gamma_L}\\
			\end{array}\right)\otimes
						\gamma^\star_{(1)}
	\end{equation}
where $e^{\pm\gamma_L}$ and $\pm\epsilon\gamma_M$ are the eigenvalues of $e^{-2\Delta\alpha_{L}P_{L}^{(0)}}$	and $-2\Delta\alpha_{M}P_{M}^{(0)}$, respectively, and $\gamma^\star_{(1)}$ is given by \eqref{eq:GammaStarI}. Thus, one can conveniently distinguish between even and odd eigenvalues.\\
One could of course just determine the eigenvalues of the matrices on both sides of \eqref{eq:TheMightyEquation} and then try to determine $\Delta\alpha_{L}$ and $\Delta\alpha_{M}$ by comparing these two sides, but there is a more efficient way of doing things i.e. taking two different traces of \eqref{eq:TheMightyEquation} in such a way that one trace picks out the purely even part and the other one the mixed even-odd part. The ordinary matrix trace used for determining $\omega_{ab}$ does the trick for the even part as can be easily seen from \eqref{eq:EWSplit}. For the mixed part we use the following ``hatted'' trace:
	\begin{equation}
		\hat{\textnormal{Tr}}\left(\mathcal{G}_a\mathcal{G}_b\right):=\left.\frac{1}{2}\frac{\extd}{\extd\epsilon}\textnormal{Tr}\left(\mathcal{G}_a\mathcal{G}_b\gamma^\star_{(2)}\right)\right|_{\epsilon=0},
	\end{equation}
where $\mathcal{G}\in\mathfrak{isl}(2,\mathbb{R})$. Using this trick one obtains the following two equations
	\begin{subequations}\label{eq:PrecurserOfTheMightyEquations}:
		\begin{align}
			2\cosh\left(\sqrt{2 c_2}\Delta\alpha_L\right)&=\left.\textnormal{Tr}\left(\Omega\right)\right|_{r_0\rightarrow\infty},\label{eq:EvenEE}\\
			2\sinh\left(\sqrt{2 c_2}\Delta\alpha_L\right)\sqrt{2\bar{c}_2}\Delta\alpha_M&=\left.\hat{\textnormal{Tr}}\left(\Omega\right)\right|_{r_0\rightarrow\infty}.\label{eq:EvenOddMixedEE}
		\end{align}
	\end{subequations}
Since the Wilson line is pushed to the boundary, $\textnormal{Tr}\left(\Omega\right)$ and thus also the left-hand side of \eqref{eq:EvenEE} will be very large and positive. As the $\cosh$ is an even function there are two branches to solve for $\Delta\alpha_L$, depending on whether $\Delta\alpha_L$ is bigger or smaller than zero. This part of the calculation is identical to the AdS$_3$ case, and thus one can use this as a pointer to choose the right branch, which in this case is
	\begin{equation}
		e^{-\sqrt{2 c_2}\Delta\alpha_L}=\left.\textnormal{Tr}\left(\Omega\right)\right|_{r_0\rightarrow\infty}.
	\end{equation}
Using this, Eq. \eqref{eq:EvenOddMixedEE} simplifies to
	\begin{equation}
		-\sqrt{2\bar{c}_2}\Delta\alpha_M\left.\textnormal{Tr}\left(\Omega\right)\right|_{r_0\rightarrow\infty}=\left.\hat{\textnormal{Tr}}\left(\Omega\right)\right|_{r_0\rightarrow\infty}.
	\end{equation}
$\Delta\alpha_L$ and $\Delta\alpha_M$ can now be determined as
		\begin{align}
		\Delta\alpha_L&=-\frac{\ln\left(\left.\textnormal{Tr}\left(\Omega\right)\right|_{r_0\rightarrow\infty}\right)}{\sqrt{2 c_2}},\\
		\Delta\alpha_M&=-\frac{\left.\hat{\textnormal{Tr}}\left(\Omega\right)\right|_{r_0\rightarrow\infty}}{\sqrt{2\bar{c}_2}\left.\textnormal{Tr}\left(\Omega\right)\right|_{r_0\rightarrow\infty}}.
	\end{align}
The entanglement entropy can thus equivalently be written as
	\begin{equation}
		S_{E}=\sqrt{2 c_2}\ln\left(\left.\hat{\textnormal{Tr}}\left(\Omega\right)\right|_{r_0\rightarrow\infty}\right)+\sqrt{2\bar{c}_2}\frac{\left.\hat{\textnormal{Tr}}\left(\Omega\right)\right|_{r_0\rightarrow\infty}}{\left.\textnormal{Tr}\left(\Omega\right)\right|_{r_0\rightarrow\infty}}.,
	\end{equation}
Writing $u_f-u_i=\Delta u$ and $\varphi_f-\varphi_i=\Delta\varphi$ then the holographic entanglement entropy for an interval with spatial extension $\Delta\varphi$ and timelike extension $\Delta u$ for the the null-orbifold ($\mathcal{M}=\mathcal{N}=0$), (global) flat space ($\mathcal{M}=-1,\mathcal{N}=0$) and FSCs ($\mathcal{M}\geq0,\mathcal{N}\neq0$), is given by
	\begin{align}
		S_{E}^{\textnormal{NO}}=&2\sqrt{2 c_2}\ln\left[\frac{r_0\Delta\phi}{2}\right]+2\sqrt{2\bar{c}_2}\frac{\Delta u}{\Delta\phi},\\
		S_{E}^{\textnormal{GFS}}=&2\sqrt{2 c_2}\ln\left[r_0\sin\left(\frac{\Delta\phi}{2}\right)\right]+\sqrt{2\bar{c}_2}\cot\left(\frac{\Delta\phi}{2}\right)\Delta u,\\
		S_{E}^{\textnormal{FSC}}=&2\sqrt{2 c_2}\ln\left[\frac{r_0\sinh\left(\frac{\sqrt{\mathcal{M}}\Delta\phi}{2}\right)}{\sqrt{\mathcal{M}}}\right]+\sqrt{2\bar{c}_2}\left(-\frac{2\mathcal{N}}{\mathcal{M}}\right.\nonumber\\
						&\left.+\sqrt{\mathcal{M}}\coth\left(\frac{\sqrt{\mathcal{M}}\Delta\phi}{2}\right)\left(\Delta u+\frac{\mathcal{N}}{\mathcal{M}}\Delta\phi\right)\right).
	\end{align}
Relating the quadratic Casimirs and central charges in a similar way as in the AdS$_3$ case, i.e., $\sqrt{2c_2}=\frac{c_L}{12}$ and $\sqrt{2\bar{c}_2}=\frac{c_M}{12}$, one obtains the following final results:
	\begin{align}
		S_{E}^{\textnormal{NO}}=&\frac{c_L}{6}\ln\left[\frac{r_0\Delta\phi}{2}\right]+\frac{c_M}{6}\frac{\Delta u}{\Delta\phi},\\
		S_{E}^{\textnormal{GFS}}=&\frac{c_L}{6}\ln\left[r_0\sin\left(\frac{\Delta\phi}{2}\right)\right]+\frac{c_M}{12}\cot\left(\frac{\Delta\phi}{2}\right)\Delta u,\\
		S_{E}^{\textnormal{FSC}}=&\frac{c_L}{6}\ln\left[\frac{r_0\sinh\left(\frac{\sqrt{\mathcal{M}}\Delta\phi}{2}\right)}{\sqrt{\mathcal{M}}}\right]+\frac{c_M}{12}\left(-\frac{2\mathcal{N}}{\mathcal{M}}\right.\nonumber\\
						&\left.+\sqrt{\mathcal{M}}\coth\left(\frac{\sqrt{\mathcal{M}}\Delta\phi}{2}\right)\left(\Delta u+\frac{\mathcal{N}}{\mathcal{M}}\Delta\phi\right)\right),
	\end{align}
which precisely coincide with the calculations done for GCFTs in the first half of our paper (where the UV cutoff $a$ is related to $r_0$ as $a=\frac{1}{r_0}$) and the results in \cite{Hosseini:2015uba} which were obtained as a limiting procedure from the AdS$_3$ results.

\subsection{Spin-3}\label{sec:HEESpin-3WilsonLine}

Having developed the flat space equivalent of the Wilson line proposal for holographic entanglement entropy in AdS$_3$ one can now also straightforwardly extend the formalism to higher-spin theories in flat space in analogy to the AdS$_3$ formulations. We illustrate how to extend our construction for the case of spin-3 flat space gravity.\\
For flat space spin-3 gravity we make the following generalizations to the ansatz \eqref{eq:SLSMFlat} used before. First we take as a gauge algebra the principal embedding of $\mathfrak{isl}(2,\mathbb{R})$ into $\mathfrak{isl}(3,\mathbb{R})$ with generators\footnote{For more details on an appropriate matrix representation see the Appendix.} $L_n,M_n,U_n,V_n$ and the following commutation relations 
	\begin{subequations}
		\begin{align}
			[L_n,\, L_m] &= (n-m) L_{n+m}  \\
			[L_n,\, M_m] &= (n-m) M_{n+m}\\
			[L_n,\, U_m] &= (2n-m) U_{n+m}  \\
			[L_n,\, V_m] &= (2n-m) V_{n+m}  \\
			[M_n,\, U_m] &= (2n-m) V_{n+m}  \\
			[U_n,\, U_m] &= (n-m)(2 n^2 + 2 m^2 - nm -8) L_{n+m} \\
			[U_n,\, V_m] &= (n-m)(2 n^2 + 2 m^2 - nm -8) M_{n+m},
		\end{align}
	\end{subequations}
where for $L_n$ and $M_n$ the indices take values $n=\pm1,0$ and for $U_m$ and $V_m$ the indices take values in $m=\pm2,\pm1,0$.\\
Next we modify the actions \eqref{eq:SLFlat} and \eqref{eq:SMFlat} in such a way that $P_L\in\left\{L_n,U_n\right\}$ and $P_M\in\left\{M_n,V_n\right\}$ and add the following constraints to the actions:
	\begin{subequations}\label{eq:FlatEOMSpin3}
		\begin{align}
		S_{L}=&\int_C\extd s\left<P_LD_LU_LU_L^{-1}\right>\nonumber\\
		&+\lambda_L\left(\left<P_L^2\right>_L-c_2\right)+\lambda^{(3)}_L\left(\left<P_L^3\right>_L-c_3\right),\\
		S_{M}=&\int_C\extd s\left<P_MU_M^{-1}D_MU_M\right>\nonumber\\
		&+\lambda_M\left(\left<P_M^2\right>_M-\bar{c}_2\right)+\lambda^{(3)}_M\left(\left<P_M^3\right>_M-\bar{c}_3\right),
		\end{align}
	\end{subequations}
where $\lambda^{(3)}_{L}$ $(\lambda^{(3)}_{M})$ are again Lagrange multipliers, $c_3$ and $\bar{c}_3$ are the cubic even and odd Casimirs\footnote{In the same sense as in the spin-2 case, i.e., $c_3$ is the $\mathfrak{sl}(3,\mathbb{R})$ Casimir and $\bar{c}_3$ the cubic Casimir of the \emph{full} $\mathfrak{isl}(3,\mathbb{R})$.}  and $\left<P_{L}^3\right>_{L}$ $(\left<P_{M}^3\right>_{M})$ is a shorthand notation for ${\left<P_{L}^3\right>_{L}=h_{abc}P_L^aP_L^bP_L^c}$ $(\left<P_{M}^3\right>_{M}=\bar{h}_{abc}P_M^aP_M^bP_M^c)$. The tensors $h_{abc}$ and $\bar{h}_{abc}$ coincide with the $\mathfrak{sl}(3,\mathbb{R})$ Killing form that defines the cubic Casimir with the only difference being that $h_{abc}$ can be obtained via 
	\begin{equation}
		\frac{1}{2}\textnormal{Tr}(G_aG_bG_c)=h_{abc},
	\end{equation}
with $G_a\in\{L_n,U_n\}$ and $\bar{h}_{abc}$ via
	\begin{equation}
		\bar{\textnormal{Tr}}(\bar{G}_a\bar{G}_b\bar{G}_c)=\bar{h}_{abc},
	\end{equation}
with $\bar{G}_a\in\{M_n,V_n\}$. 
The EOMs of \eqref{eq:FlatEOMSpin3} are given by
	\begin{gather}
	D_{L}U_{L}U_{L}^{-1}+2\lambda_{L}P_{L}+3\lambda^{(3)}_{L}P_{L}\times P_{L}=0,\nonumber\\
	\frac{\extd}{\extd s}P_{L}=0,\quad (L\leftrightarrow M)
	\end{gather}
in addition to the constraints $\left< P_L^2\right>_L=c_2$, $\left< P_M^2\right>_M=\bar{c}_2$, $\left< P_L^3\right>_L=c_3$ and $\left< P_M^3\right>_M=\bar{c}_3$. $P_{L}\times P_{L}=0$ and $P_{M}\times P_{M}=0$ are shorthand notations for $P_{L}\times P_{L}=h_{abc}G^aP_L^bP_L^c$ and $P_{M}\times P_{M}=\bar{h}_{abc}\bar{G}^aP_M^bP_M^c$. Using again the ``nothingness'' trick one obtains the following solution for $A_L=A_M=0$:
	\begin{gather}\label{eq:U0SolutionSpin3}
		U^{(0)}_{L}=u_{L}^{(0)}e^{\left(-2\alpha_{L}(s)P_{L}^{(0)}-3\alpha^{(3)}_{L}(s)P_{L}^{(0)}\times P_{L}^{(0)}\right)},\non \\ \frac{d\alpha_{L}(s)}{ds}=\lambda_{L}(s),\quad \frac{d\alpha^{(3)}_{L}(s)}{ds}=\lambda^{(3)}_{L}(s),\quad (L\leftrightarrow M).
	\end{gather}
The on-shell action is given by
	\begin{align}
		S_{L}^{\textnormal{on-shell}}=&-2\Delta\alpha_{L}c_2-3\Delta\alpha^{(3)}_{L}c_3,\nonumber\\ 
		S_{M}^{\textnormal{on-shell}}=&-2\Delta\alpha_{M}\bar{c}_2-3\Delta\alpha^{(3)}_{M}\bar{c}_3.
	\end{align}
Now one can define
	\begin{align}
		\mathbb{P}_{L}:=&-2\Delta\alpha_{L}(s)P_{L}^{(0)}-3\Delta\alpha^{(3)}_{L}(s)P_{L}^{(0)}\times P_{L}^{(0)},\nonumber\\
		\mathbb{P}_{M}:=&-2\Delta\alpha_{M}(s)P_{M}^{(0)}-3\Delta\alpha^{(3)}_{M}(s)P_{M}^{(0)}\times P_{M}^{(0)},
	\end{align}
and perform the same steps as in the spin-2 case in order to obtain the spin-3 analogue of \eqref{eq:TheMightyEquation},
	\begin{equation}
		e^{\mathbb{P}_L+\mathbb{P}_M}=\Omega,
	\end{equation}
where $\Omega$ is the same expression as in \eqref{eq:TheMightyEquation} with the exception that $U$ now takes values in $\mathfrak{isl}(3,\mathbb{R})$ and $A_{i/f}$ are determined by the corresponding spin-3 Chern-Simons connection.\\
Using the EOMs one can further simplify this equation to the following set of equations:
	\begin{align}
		-2\Delta\alpha_{L}c_2-3\Delta\alpha^{(3)}_{L}c_3=&\frac{1}{2}\textnormal{Tr}\left[\ln\left(\Omega\right)P_{L}^{(0)}\right]=S_{L}^{\textnormal{on-shell}},\\
		-2\Delta\alpha_{M}\bar{c}_2-3\Delta\alpha^{(3)}_{M}\bar{c}_3=&\bar{\textnormal{Tr}}\left[\ln\left(\Omega\right)P_{M}^{(0)}\right]=S_{M}^{\textnormal{on-shell}}.
	\end{align}
Since in the semiclassical limit the entanglement entropy is proportional to the on-shell action, one can thus write the entanglement entropy as
	\begin{align}
		S_{E}=&S_{L}^{\textnormal{on-shell}}+S_{M}^{\textnormal{on-shell}}\nonumber\\
		=&\frac{1}{2}\textnormal{Tr}\left[\ln\left(\Omega\right)P_{L}^{(0)}\right]+\bar{\textnormal{Tr}}\left[\ln\left(\Omega\right)P_{M}^{(0)}\right],
	\end{align}
or equivalently as
	\begin{equation}
		S_{E}=-2\Delta\alpha_{L}c_2-3\Delta\alpha^{(3)}_{L}c_3-2\Delta\alpha_{M}\bar{c}_2-3\Delta\alpha^{(3)}_{M}\bar{c}_3.
	\end{equation}
One can now use this expression and the connection
	\begin{equation}
		\mathcal{A}=b^{-1}\extd b+b^{-1}ab,\quad b=e^{\frac{r}{2} M_{-1}},
	\end{equation}
with	
	\begin{align}
		a=&\left( M_{1}-\frac{\mathcal{M}}{4}M_{-1}-\frac{\mathcal{V}}{16}V_{-2}\right)\extd u\nonumber\\
		&+\left(L_{1}-\frac{\mathcal{M}}{4}L_{-1}-\frac{\mathcal{N}}{2}M_{-1}-\frac{\mathcal{V}}{16}U_{-2}-\frac{\mathcal{Z}}{8}V_{-2}\right)\extd\phi,		
	\end{align}
where $\mathcal{V}$ and $\mathcal{Z}$ are the even and odd spin-3 charges, respectively, and determine the holographic entanglement entropy of a spin-3 charged FSC along the same lines as in the spin-2 case in the previous subsection. Since the results are rather lengthy we will, however, not display them here explicitly.
		
\section{Thermal Entropy of Flat Space Cosmologies}\label{sec:ThermalEntropyFSC}

\subsection{Spin-2}

In this section we show how to use Wilson lines to determine the thermal entropy of FSCs. In order to do this one has to consider a closed Wilson loop around the noncontractable cycle of the FSC, i.e., the $\phi$ cycle. Since one is now dealing with a Wilson \emph{loop} instead of a Wilson \emph{line}, the topological probe should be continuous at the initial and final points and thus periodic, i.e.,
	\begin{equation}
		U_i=U_f,\quad\textnormal{and}\quad P_i=P_f.
	\end{equation}
Since $P(s)=AP^{(0)}A^{-1}$ these boundary conditions imply that
	\begin{equation}
		\left[P^{(0)},A_i^{-1}A_f\right]=\left[P^{(0)}_L+P^{(0)}_M,A_i^{-1}A_f\right]=0.
	\end{equation}
$P^{(0)}_L$ and $P^{(0)}_M$ commute with each other and thus one can simultaneously diagonalize them. This in turn means that $\left[P^{(0)}_L,A_i^{-1}A_f\right]$ and $\left[P^{(0)}_M,A_i^{-1}A_f\right]$ vanish simultaneously. For the topological probe one finds the following relation
	\begin{equation}\label{eq:ThermalDefEq}	
	e^{-2\Lambda}=\left(u^{(0)}_Lu^{(0)}_M\right)^{-1}\left(A^{-1}_iA_f\right)^{-1}u^{(0)}_Lu^{(0)}_MA^{-1}_iA_f,
	\end{equation}
where ${\Lambda=\Delta\alpha_LP^{(0)}_L+\Delta\alpha_MP^{(0)}_M}$ and assuming that $P^{(0)}_L$, $P^{(0)}_M$ and $A^{-1}_iA_f$ have already been diagonalized. Since the noncontractible cycle is the $\phi$ cycle, $A^{-1}_iA_f$ reduces to the holonomy around this cycle, i.e.,
	\begin{equation}
		A^{-1}_iA_f=e^{-2\pi \lambda_\phi},
	\end{equation}
where $\lambda_\phi$ denotes the diagonalized form of $a_\phi$ given in \eqref{eq:Spin2ChernSimonsConnection}. Since the right-hand side of \eqref{eq:ThermalDefEq} should be nontrivial one has to choose $u^{(0)}_Lu^{(0)}_M$ in such a way that
	\begin{equation}
		\left(u^{(0)}_Lu^{(0)}_M\right)^{-1}\left(A^{-1}_iA_f\right)^{-1}u^{(0)}_Lu^{(0)}_M=A^{-1}_iA_f,
	\end{equation}
or equivalently
	\begin{equation}\label{eq:ThermalEntropyULUMAssumption}
		\left(u^{(0)}_Lu^{(0)}_M\right)^{-1}e^{2\pi \lambda_\phi}u^{(0)}_Lu^{(0)}_M=e^{-2\pi \lambda_\phi}.
	\end{equation}
Using \eqref{eq:ThermalEntropyULUMAssumption} one finds that the following equation has to be satisfied
	\begin{equation}\label{eq:ThermalEquation}
		\Delta\alpha_LP^{(0)}_L+\Delta\alpha_MP^{(0)}_M=2\pi\lambda_\phi.
	\end{equation}
Since $P^{(0)}_L$ and $P^{(0)}_M$ are traceless and the constraints fix $\frac{1}{2}\textnormal{Tr}\left(P_L^2\right)=c_2$ and $\bar{\textnormal{Tr}}\left(P_M^2\right)=\bar{c}_2$[and $\frac{1}{2}\textnormal{Tr}\left(P_LM_0\right)=0$, $\bar{\textnormal{Tr}}\left(P_ML_0\right)=0$] one immediately sees that the eigenvalues of $P^{(0)}_L$ and $P^{(0)}_M$ are $\pm\sqrt{\frac{c_2}{2}}$ and $\pm\epsilon\sqrt{\frac{\bar{c}_2}{2}}$ in each of the $\mathfrak{sl}(2,\mathbb{R})$ blocks which were employed in the construction of the $\mathfrak{isl}(2,\mathbb{R})$ matrix representation found in appendix \ref{sec:App}. Phrased in terms of traces this means that
	\begin{equation}
		\frac{1}{2}\textnormal{Tr}\left(P_LL_0\right)=\sqrt{\frac{c_2}{2}},\quad\textnormal{and}\quad\bar{\textnormal{Tr}}\left(P_MM_0\right)=\sqrt{\frac{\bar{c}_2}{2}}.
	\end{equation}
Thus, by multiplying \eqref{eq:ThermalEquation} with $L_0$ and $M_0$ and taking either the trace or twisted trace, one can determine $\Delta\alpha_L$ and $\Delta\alpha_M$ via
	\begin{align}
		\Delta\alpha_L=&\pi\sqrt{\frac{2}{c_2}}\textnormal{Tr}\left(\lambda_\phi L_0\right),\nonumber\\
		\Delta\alpha_M=&2\pi\sqrt{\frac{2}{\bar{c}_2}}\bar{\textnormal{Tr}}\left(\lambda_\phi M_0\right).
	\end{align}
For the FSC given by the connection \eqref{eq:Spin2ChernSimonsConnection} one obtains the following values for $\Delta\alpha_{L}$ and $\Delta\alpha_{M}$
	\begin{equation}
		\Delta\alpha_L=-\pi\sqrt{\frac{2\mathcal{M}}{c_2}},\quad\textnormal{and}\quad\Delta\alpha_M=-\frac{\pi\mathcal{N}}{\sqrt{\mathcal{M}}}\sqrt{\frac{2}{\bar{c}_2}}.
	\end{equation}
Making, in addition, the same identifications of the quadratic Casimirs as in the case of the thermal entropy, this yields the following thermal entropy
	\begin{equation}\label{eq:FSCThermEnt}
		S_{Therm}=\frac{\pi}{6}\left(c_L\sqrt{\mathcal{M}}+c_M\frac{\mathcal{N}}{\sqrt{\mathcal{M}}}\right).
	\end{equation}
Taking into account 
	\begin{equation}
		\mathcal{M}=\frac{24h_M}{c_M},\quad\mathcal{N}=\frac{12\left(c_Mh_L-c_Lh_M\right)}{c_M^2},
	\end{equation}
one can immediately check that this is exactly the same result as \eqref{eq:PartGCAEntropy} obtained previously in Sec. \ref{sec:ThermalEntropy} or alternatively by performing an \.In\"on\"u--Wigner contraction of the inner horizon thermal entropy of the BTZ black hole in AdS$_3$ \cite{Riegler:2014bia}.
				
\subsection{Spin-3}

As in the spin-2 case we now determine the thermal entropy for a spin-3 charged FSC holographically. In order to proceed one has to perform the same steps as in the spin-2 case i.e. using a closed Wilson loop around the $\phi$ direction. With the notation used in Sec. \ref{sec:HEESpin-3WilsonLine} this leads to the following equation which has to be solved
	\begin{equation}
		\mathbb{P}_L+\mathbb{P}_M=2\pi\lambda^{(3)}_\phi,
	\end{equation}
where $\lambda_\phi^{(3)}$ denotes the diagonalized form of $a_\phi^{(3)}$ which is given by \cite{Gary:2014ppa} 
	\begin{equation}
		a_\phi^{(3)}=L_1-\frac{\mathcal{M}}{4}L_{-1}+\frac{\mathcal{V}}{2}U_{-2}-\frac{\mathcal{N}}{2}M_{-1}+\mathcal{Z}V_{-2},
 	\end{equation}
and the eigenvalues of $a_\phi^{(3)}$ are ordered in such a way that they coincide with the spin-2 case for vanishing spin-3 charges $\mathcal{V}$ and $\mathcal{Z}$. Setting\footnote{This choice of $P_{L}^{(0)}$ and  $P_{M}^{(0)}$ is tantamount to setting the cubic Casimirs $c_3,\bar{c}_3$ to zero.} 
	\begin{equation}
		P_L^{(0)}=\sqrt{\frac{c_L}{2}}L_0,\quad P_M^{(0)}=\sqrt{\frac{c_M}{2}}M_0,
	\end{equation}
this equation simplifies to
	\begin{align}\label{eq:Spin3DiagMagicEq}		
	2\pi\lambda^{(3)}_\phi=&-\sqrt{2c_2}\Delta\alpha_LL_0-2c_2\Delta\alpha_L^{(3)}U_0\nonumber\\
	&-\sqrt{2\bar{c}_2}\Delta\alpha_MM_0-2\bar{c}_2\Delta\alpha_M^{(3)}V_0.
	\end{align}
As in the spin-2 case one can now solve for $\Delta\alpha_L$ and $\Delta\alpha_M$ by multiplying either $L_0$ or $M_0$ on both sides of \eqref{eq:Spin3DiagMagicEq} and taking the (twisted) trace. This yields the following relations:
	\begin{align}
		\Delta\alpha_L=&-\frac{\pi}{2\sqrt{2c_2}}\textnormal{Tr}\left(\lambda_\phi^{(3)}L_0\right),\nonumber\\
		\Delta\alpha_M=&-\frac{\pi}{\sqrt{2\bar{c}_2}}\bar{\textnormal{Tr}}\left(\lambda_\phi^{(3)}M_0\right).
	\end{align}
Thus one can write the thermal entropy for the spin-3 charged FSC as
	\begin{equation}\label{eq:GeneralThermSpin3Entropy}
		S_{\textnormal{Th}}=\pi\left(\sqrt{\frac{c_2}{2}}\textnormal{Tr}\left(\lambda_\phi^{(3)}L_0\right)+\sqrt{2\bar{c}_2}\bar{\textnormal{Tr}}\left(\lambda_\phi^{(3)}M_0\right)\right).
	\end{equation}
Replacing again the quadratic Casimirs in the same fashion as in the section before and evaluating the traces, one obtains the following expression:
	\begin{align}\label{eq:ThermSpin3Entropy}
		S_{\textnormal{Th}}=&\frac{\pi}{6}\left(c_L\sqrt{\mathcal{M}}\sqrt{1-\frac{3}{4\mathcal{R}}}\right.\nonumber\\
		&\left.+c_M\frac{\mathcal{N}}{\sqrt{\mathcal{M}}}\frac{\left(2\mathcal{R}-3+12\mathcal{P}\sqrt{\mathcal{R}}\right)}{2\left(\mathcal{R}-3\right)\sqrt{1-\frac{3}{4\mathcal{R}}}}\right),
	\end{align}
where we have rewritten  $\mathcal{V}$ and $\mathcal{Z}$ in terms of the dimensionless parameters $\mathcal{R}$ and $\mathcal{P}$ as
	\begin{equation}
		\frac{\mathcal{|V|}}{\mathcal{M}^{\frac{3}{2}}}=\frac{\mathcal{R}-1}{4\mathcal{R}^\frac{3}{2}},\quad \frac{\mathcal{Z}}{\mathcal{N}\sqrt{\mathcal{M}}}=\mathcal{P},
	\end{equation}
which exactly coincides with the results obtained in \cite{Riegler:2014bia,Gary:2014ppa}.	

\section{Conclusion and Outlook}

In this paper we gave a detailed description of how to determine entanglement entropy and thermal entropy of FSCs holographically for flat space in $2+1$ dimensions using Wilson lines, and we, compared our results with field theoretic calculations employing GCFT techniques. We also successfully extended the holographic description to incorporate higher-spin symmetries and determine the thermal entropy of FSCs with spin-2 and spin-3 charges.\\
Now having adapted the Wilson line approach to holographic entanglement entropy for use in flat space--including higher-spin symmetries--one possible further application of the methods presented in this paper would be to determine both entanglement entropy and thermal entropy for other flat space scenarios such as flat space in Rindler coordinates.\\
Another crucial aspect for a better understanding of flat space higher-spin holography would be to extend the GCFT calculations done in this paper to incorporate also higher-spin fields. This would give another cross-check of the holographic spin-3 results presented in this paper.\\
For flat space Einstein gravity one would expect that one could equivalently derive entanglement entropy holographically by using geometric quantities such as geodesics or generalized constructions as in \cite{Hubeny:2007xt}. Thus, it would be very interesting to relate such a geometric derivation of holographic entanglement entropy with the construction using Wilson lines as employed in this paper.\\
From a GCFT point of view it would also be of great interest to investigate flat space analogues of the Cardy-Calabrese quench \cite{Calabrese:2007rg} and its gravity dual. Another entanglement related question of interest from a GCFT perspective would be whether or not one can find a GCFT analogue of the CFT construction on entanglement negativity \cite{Calabrese:2012ew} and again a possible bulk dual similar to, e.g., Ref. \cite{Perlmutter:2015vma}.\\
Further interesting extensions would also be various (holographic) checks of information theoretic properties like, for example, strong subadditivity and monogamy of mutual information for flat space holography involving GCFTs.\\
Finally, it would be very interesting to incorporate the proof of the Ryu-Takayanagi proposal by Lewkowycz and Maldacena \cite{Lewkowycz:2013nqa} and possibly adapt it to flat space.
\subsection*{Acknowledgments}

We are grateful to Daniel Grumiller and Arjun Bagchi for collaboration at the initial stage of this project and for many helpful insights. In addition we want to thank Stephane Detournay, Shankhadeep Chakrabortty and Alejandra Castro for useful discussions. The research of RB was supported by INSPIRE Faculty award of the Department
of Science and Technology, India. This research of MR was supported by the FWF Project No. P27182-N27 and the START Project No. Y435-N16. MR was also supported by a \emph{DOC} fellowship of the Austrian Academy of Sciences, the \emph{Doktoratskolleg Particles and Interactions} (FWF Project No. DKW1252-N27) and the FWF Project No. I 1030-N27.
				
\begin{appendix}

\section{Matrix Representations}\label{sec:App}
Throughout this paper we use the following matrix representations of $\mathfrak{isl}(2,\mathbb{R})$ and $\mathfrak{isl}(3,\mathbb{R})$ generators in terms of $4\times4$ and $6\times6$ block-diagonal matrices. This block structure is a remnant of the decomposition of the AdS$_3$ symmetry algebra $\mathfrak{so}(2,2)\sim\mathfrak{sl}(2,\mathbb{R})\oplus\mathfrak{sl}(2,\mathbb{R})$ before the \.In\"on\"u--Wigner contraction. In the following expressions $\epsilon$ denotes the Grassmann parameter first introduced in \cite{Krishnan:2013wta}.\\
\subsection{$\mathfrak{isl}(2,\mathbb{R})$}	
	\begin{align}
		L_1=&\left(
			\begin{array}{cccc}
				 0 & 0 & 0 & 0\\
				 1 & 0 & 0 & 0\\
				 0 & 0 & 0 & 0\\
				 0 & 0 & 1 & 0\\
			\end{array}\right),\quad
		L_0=\frac{1}{2}\left(
			\begin{array}{cccc}
				 1 & 0 & 0 & 0\\
				 0 & -1 & 0 & 0\\
				 0 & 0 & 1 & 0\\
				 0 & 0 & 0 & -1\\
			\end{array}\right),\nonumber\\
		L_{-1}=&\left(
			\begin{array}{cccc}
				 0 & -1 & 0 & 0\\
				 0 & 0 & 0 & 0\\
				 0 & 0 & 0 & -1\\
				 0 & 0 & 0 & 0\\
			\end{array}\right),\quad
		M_1=\left(
			\begin{array}{cccc}
				 0 & 0 & 0 & 0\\
				 \epsilon & 0 & 0 & 0\\
				 0 & 0 & 0 & 0\\
				 0 & 0 & -\epsilon & 0\\
			\end{array}\right),\nonumber\\
		M_0=&\frac{1}{2}\left(
			\begin{array}{cccc}
				 \epsilon & 0 & 0 & 0\\
				 0 & -\epsilon & 0 & 0\\
				 0 & 0 & -\epsilon & 0\\
				 0 & 0 & 0 & \epsilon\\
			\end{array}\right),\quad
		M_{-1}=\left(
			\begin{array}{cccc}
				 0 & -\epsilon & 0 & 0\\
				 0 & 0 & 0 & 0\\
				 0 & 0 & 0 & \epsilon\\
				 0 & 0 & 0 & 0\\
			\end{array}\right).
	\end{align}

\subsection{$\mathfrak{isl}(3,\mathbb{R})$}
Even spin-2 generators
	\begin{align}
		L_1=&\left(
			\begin{array}{ccc}
				 0 & 0 & 0 \\
				 1 & 0 & 0 \\
				 0 & 1 & 0 \\
			\end{array}\right)\otimes\unity_{2\times2},\quad
		L_0=\left(
			\begin{array}{ccc}
				 1 & 0 & 0 \\
				 0 & 0 & 0 \\
				 0 & 0 & -1 \\
			\end{array}\right)\otimes\unity_{2\times2},\nonumber\\
		L_{-1}=&\left(
			\begin{array}{ccc}
				 0 & -2 & 0 \\
				 0 & 0 & -2 \\
				 0 & 0 & 0 \\
			\end{array}\right)\otimes\unity_{2\times2}.
	\end{align}
Even spin-3 generators
	\begin{align}
		U_2=&\left(
			\begin{array}{ccc}
				 0 & 0 & 0 \\
				 0 & 0 & 0 \\
				 2 & 0 & 0 \\
			\end{array}\right)\otimes\unity_{2\times2},\quad
		U_1=\left(
			\begin{array}{ccc}
				 0 & 0 & 0 \\
				 1 & 0 & 0 \\
				 0 & -1 & 0 \\
			\end{array}\right)\otimes\unity_{2\times2},\nonumber\\
		U_{0}=&\left(
			\begin{array}{ccc}
				 \frac{2}{3} & 0 & 0 \\
				 0 & -\frac{4}{3} & 0 \\
				 0 & 0 & \frac{2}{3} \\
			\end{array}\right)\otimes\unity_{2\times2},\quad
		U_{-1}=\left(
			\begin{array}{ccc}
				 0 & -2 & 0 \\
				 0 & 0 & 2 \\
				 0 & 0 & 0 \\
			\end{array}\right)\otimes\unity_{2\times2},\nonumber\\
		U_{-2}=&\left(
			\begin{array}{ccc}
				 0 & 0 & 8 \\
				 0 & 0 & 0 \\
				 0 & 0 & 0 \\
			\end{array}\right)\otimes\unity_{2\times2}.
	\end{align}
All odd generators can be obtained as a matrix product of the corresponding even generators and $\gamma^\star_{(3)}$ as defined in \eqref{eq:GammaStarI} i.e.
	\begin{equation}
		M_n=\epsilon L_n\times\gamma^\star_{(3)},\quad V_n=\epsilon U_n\times\gamma^\star_{(3)}.
	\end{equation} 
\end{appendix}

\providecommand{\href}[2]{#2}\begingroup\raggedright\endgroup


\begin{thebibliography}{10}

\bibitem{Bagchi:2014iea}
A.~Bagchi, R.~Basu, D.~Grumiller, and M.~Riegler, ``{Entanglement entropy in
  Galilean conformal field theories and flat holography},'' {\em Phys. Rev.
  Lett.} {\bf 114} (2015), no.~11, 111602,
\href{http://www.arXiv.org/abs/1410.4089}{{\tt 1410.4089}}.

\bibitem{Holzhey:1994we}
C.~Holzhey, F.~Larsen, and F.~Wilczek, ``{Geometric and renormalized entropy in
  conformal field theory},'' {\em Nucl. Phys.} {\bf B424} (1994) 443--467,
\href{http://www.arXiv.org/abs/hep-th/9403108}{{\tt hep-th/9403108}}.

\bibitem{Calabrese:2004eu}
P.~Calabrese and J.~L. Cardy, ``{Entanglement entropy and quantum field
  theory},'' {\em J. Stat. Mech.} {\bf 0406} (2004) P06002,
\href{http://www.arXiv.org/abs/hep-th/0405152}{{\tt hep-th/0405152}}.

\bibitem{Susskind:1994vu}
L.~Susskind, ``{The World as a hologram},'' {\em J. Math. Phys.} {\bf 36}
  (1995) 6377--6396,
\href{http://www.arXiv.org/abs/hep-th/9409089}{{\tt hep-th/9409089}}.

\bibitem{Maldacena:1997re}
J.~M. Maldacena, ``{The Large N limit of superconformal field theories and
  supergravity},'' {\em Int. J. Theor. Phys.} {\bf 38} (1999) 1113--1133,
  \href{http://www.arXiv.org/abs/hep-th/9711200}{{\tt hep-th/9711200}}.
[Adv. Theor. Math. Phys.2,231(1998)].

\bibitem{Ryu:2006bv}
S.~Ryu and T.~Takayanagi, ``{Holographic derivation of entanglement entropy
  from AdS/CFT},'' {\em Phys. Rev. Lett.} {\bf 96} (2006) 181602,
\href{http://www.arXiv.org/abs/hep-th/0603001}{{\tt hep-th/0603001}}.

\bibitem{Bombelli:1986rw}
L.~Bombelli, R.~K. Koul, J.~Lee, and R.~D. Sorkin, ``{A Quantum Source of
  Entropy for Black Holes},'' {\em Phys. Rev.} {\bf D34} (1986)
373--383.

\bibitem{Srednicki:1993im}
M.~Srednicki, ``{Entropy and area},'' {\em Phys. Rev. Lett.} {\bf 71} (1993)
  666--669,
\href{http://www.arXiv.org/abs/hep-th/9303048}{{\tt hep-th/9303048}}.

\bibitem{Kitaev:2005dm}
A.~Kitaev and J.~Preskill, ``{Topological entanglement entropy},'' {\em Phys.
  Rev. Lett.} {\bf 96} (2006) 110404,
\href{http://www.arXiv.org/abs/hep-th/0510092}{{\tt hep-th/0510092}}.

\bibitem{Ryu:2006ef}
S.~Ryu and T.~Takayanagi, ``{Aspects of Holographic Entanglement Entropy},''
  {\em JHEP} {\bf 08} (2006) 045,
\href{http://www.arXiv.org/abs/hep-th/0605073}{{\tt hep-th/0605073}}.

\bibitem{Solodukhin:2006xv}
S.~N. Solodukhin, ``{Entanglement entropy of black holes and AdS/CFT
  correspondence},'' {\em Phys. Rev. Lett.} {\bf 97} (2006) 201601,
\href{http://www.arXiv.org/abs/hep-th/0606205}{{\tt hep-th/0606205}}.

\bibitem{Hubeny:2007xt}
V.~E. Hubeny, M.~Rangamani, and T.~Takayanagi, ``{A Covariant holographic
  entanglement entropy proposal},'' {\em JHEP} {\bf 07} (2007) 062,
\href{http://www.arXiv.org/abs/0705.0016}{{\tt 0705.0016}}.

\bibitem{Nishioka:2009un}
T.~Nishioka, S.~Ryu, and T.~Takayanagi, ``{Holographic Entanglement Entropy: An
  Overview},'' {\em J. Phys.} {\bf A42} (2009) 504008,
\href{http://www.arXiv.org/abs/0905.0932}{{\tt 0905.0932}}.

\bibitem{Headrick:2010zt}
M.~Headrick, ``{Entanglement Renyi entropies in holographic theories},'' {\em
  Phys. Rev.} {\bf D82} (2010) 126010,
\href{http://www.arXiv.org/abs/1006.0047}{{\tt 1006.0047}}.

\bibitem{Casini:2011kv}
H.~Casini, M.~Huerta, and R.~C. Myers, ``{Towards a derivation of holographic
  entanglement entropy},'' {\em JHEP} {\bf 05} (2011) 036,
\href{http://www.arXiv.org/abs/1102.0440}{{\tt 1102.0440}}.

\bibitem{Almheiri:2012rt}
A.~Almheiri, D.~Marolf, J.~Polchinski, and J.~Sully, ``{Black Holes:
  Complementarity or Firewalls?},'' {\em JHEP} {\bf 02} (2013) 062,
\href{http://www.arXiv.org/abs/1207.3123}{{\tt 1207.3123}}.

\bibitem{Maldacena:2013xja}
J.~Maldacena and L.~Susskind, ``{Cool horizons for entangled black holes},''
  {\em Fortsch. Phys.} {\bf 61} (2013) 781--811,
\href{http://www.arXiv.org/abs/1306.0533}{{\tt 1306.0533}}.

\bibitem{Marolf:2013dba}
D.~Marolf and J.~Polchinski, ``{Gauge/Gravity Duality and the Black Hole
  Interior},'' {\em Phys. Rev. Lett.} {\bf 111} (2013) 171301,
\href{http://www.arXiv.org/abs/1307.4706}{{\tt 1307.4706}}.

\bibitem{Papadodimas:2013wnh}
K.~Papadodimas and S.~Raju, ``{Black Hole Interior in the Holographic
  Correspondence and the Information Paradox},'' {\em Phys. Rev. Lett.} {\bf
  112} (2014), no.~5, 051301,
\href{http://www.arXiv.org/abs/1310.6334}{{\tt 1310.6334}}.

\bibitem{Nozaki:2013wia}
M.~Nozaki, T.~Numasawa, and T.~Takayanagi, ``{Holographic Local Quenches and
  Entanglement Density},'' {\em JHEP} {\bf 05} (2013) 080,
\href{http://www.arXiv.org/abs/1302.5703}{{\tt 1302.5703}}.

\bibitem{Harlow:2014yka}
D.~Harlow, ``{Jerusalem Lectures on Black Holes and Quantum Information},''
\href{http://www.arXiv.org/abs/1409.1231}{{\tt 1409.1231}}.

\bibitem{Ecker:2015kna}
C.~Ecker, D.~Grumiller, and S.~A. Stricker, ``{Evolution of holographic
  entanglement entropy in an anisotropic system},'' {\em JHEP} {\bf 07} (2015)
  146,
\href{http://www.arXiv.org/abs/1506.02658}{{\tt 1506.02658}}.

\bibitem{Klebanov:2002ja}
I.~R. Klebanov and A.~M. Polyakov, ``{AdS dual of the critical O(N) vector
  model},'' {\em Phys. Lett.} {\bf B550} (2002) 213--219,
\href{http://www.arXiv.org/abs/hep-th/0210114}{{\tt hep-th/0210114}}.

\bibitem{Mikhailov:2002bp}
A.~Mikhailov, ``{Notes on higher spin symmetries},''
\href{http://www.arXiv.org/abs/hep-th/0201019}{{\tt hep-th/0201019}}.

\bibitem{Sezgin:2002rt}
E.~Sezgin and P.~Sundell, ``{Massless higher spins and holography},'' {\em
  Nucl. Phys.} {\bf B644} (2002) 303--370,
  \href{http://www.arXiv.org/abs/hep-th/0205131}{{\tt hep-th/0205131}}.
[Erratum: Nucl. Phys.B660,403(2003)].

\bibitem{Maldacena:2011jn}
J.~Maldacena and A.~Zhiboedov, ``{Constraining Conformal Field Theories with A
  Higher Spin Symmetry},'' {\em J. Phys.} {\bf A46} (2013) 214011,
\href{http://www.arXiv.org/abs/1112.1016}{{\tt 1112.1016}}.

\bibitem{Maldacena:2012sf}
J.~Maldacena and A.~Zhiboedov, ``{Constraining conformal field theories with a
  slightly broken higher spin symmetry},'' {\em Class. Quant. Grav.} {\bf 30}
  (2013) 104003,
\href{http://www.arXiv.org/abs/1204.3882}{{\tt 1204.3882}}.

\bibitem{Gubser:1998bc}
S.~S. Gubser, I.~R. Klebanov, and A.~M. Polyakov, ``{Gauge theory correlators
  from noncritical string theory},'' {\em Phys. Lett.} {\bf B428} (1998)
  105--114,
\href{http://www.arXiv.org/abs/hep-th/9802109}{{\tt hep-th/9802109}}.

\bibitem{Witten:1998qj}
E.~Witten, ``{Anti-de Sitter space and holography},'' {\em Adv. Theor. Math.
  Phys.} {\bf 2} (1998) 253--291,
\href{http://www.arXiv.org/abs/hep-th/9802150}{{\tt hep-th/9802150}}.

\bibitem{Gaberdiel:2010pz}
M.~R. Gaberdiel and R.~Gopakumar, ``{An AdS$_3$ Dual for Minimal Model CFTs},''
  {\em Phys. Rev.} {\bf D83} (2011) 066007,
\href{http://www.arXiv.org/abs/1011.2986}{{\tt 1011.2986}}.

\bibitem{Ammon:2012wc}
M.~Ammon, M.~Gutperle, P.~Kraus, and E.~Perlmutter, ``{Black holes in three
  dimensional higher spin gravity: A review},'' {\em J. Phys.} {\bf A46} (2013)
  214001,
\href{http://www.arXiv.org/abs/1208.5182}{{\tt 1208.5182}}.

\bibitem{Gaberdiel:2012uj}
M.~R. Gaberdiel and R.~Gopakumar, ``{Minimal Model Holography},'' {\em J.
  Phys.} {\bf A46} (2013) 214002,
\href{http://www.arXiv.org/abs/1207.6697}{{\tt 1207.6697}}.

\bibitem{Perez:2014pya}
A.~Perez, D.~Tempo, and R.~Troncoso, ``{Higher Spin Black Holes},'' {\em Lect.
  Notes Phys.} {\bf 892} (2015) 265--288,
\href{http://www.arXiv.org/abs/1402.1465}{{\tt 1402.1465}}.

\bibitem{Afshar:2014rwa}
H.~Afshar, A.~Bagchi, S.~Detournay, D.~Grumiller, S.~Prohazka, and M.~Riegler,
  ``{Holographic Chern-Simons Theories},'' {\em Lect. Notes Phys.} {\bf 892}
  (2015) 311--329,
\href{http://www.arXiv.org/abs/1404.1919}{{\tt 1404.1919}}.

\bibitem{Gary:2012ms}
M.~Gary, D.~Grumiller, and R.~Rashkov, ``{Towards non-AdS holography in
  3-dimensional higher spin gravity},'' {\em JHEP} {\bf 03} (2012) 022,
\href{http://www.arXiv.org/abs/1201.0013}{{\tt 1201.0013}}.

\bibitem{Afshar:2012nk}
H.~Afshar, M.~Gary, D.~Grumiller, R.~Rashkov, and M.~Riegler, ``{Non-AdS
  holography in 3-dimensional higher spin gravity - General recipe and
  example},'' {\em JHEP} {\bf 11} (2012) 099,
\href{http://www.arXiv.org/abs/1209.2860}{{\tt 1209.2860}}.

\bibitem{Afshar:2012hc}
H.~Afshar, M.~Gary, D.~Grumiller, R.~Rashkov, and M.~Riegler, ``{Semi-classical
  unitarity in 3-dimensional higher-spin gravity for non-principal
  embeddings},'' {\em Class. Quant. Grav.} {\bf 30} (2013) 104004,
\href{http://www.arXiv.org/abs/1211.4454}{{\tt 1211.4454}}.

\bibitem{Gutperle:2013oxa}
M.~Gutperle, E.~Hijano, and J.~Samani, ``{Lifshitz black holes in higher spin
  gravity},'' {\em JHEP} {\bf 04} (2014) 020,
\href{http://www.arXiv.org/abs/1310.0837}{{\tt 1310.0837}}.

\bibitem{Gary:2014mca}
M.~Gary, D.~Grumiller, S.~Prohazka, and S.-J. Rey, ``{Lifshitz Holography with
  Isotropic Scale Invariance},'' {\em JHEP} {\bf 08} (2014) 001,
\href{http://www.arXiv.org/abs/1406.1468}{{\tt 1406.1468}}.

\bibitem{Krishnan:2013zya}
C.~Krishnan, A.~Raju, S.~Roy, and S.~Thakur, ``{Higher Spin Cosmology},'' {\em
  Phys. Rev.} {\bf D89} (2014), no.~4, 045007,
\href{http://www.arXiv.org/abs/1308.6741}{{\tt 1308.6741}}.

\bibitem{Afshar:2013vka}
H.~Afshar, A.~Bagchi, R.~Fareghbal, D.~Grumiller, and J.~Rosseel, ``{Spin-3
  Gravity in Three-Dimensional Flat Space},'' {\em Phys. Rev. Lett.} {\bf 111}
  (2013), no.~12, 121603,
\href{http://www.arXiv.org/abs/1307.4768}{{\tt 1307.4768}}.

\bibitem{Gonzalez:2013oaa}
H.~A. Gonzalez, J.~Matulich, M.~Pino, and R.~Troncoso, ``{Asymptotically flat
  spacetimes in three-dimensional higher spin gravity},'' {\em JHEP} {\bf 09}
  (2013) 016,
\href{http://www.arXiv.org/abs/1307.5651}{{\tt 1307.5651}}.

\bibitem{Coleman:1967ad}
S.~R. Coleman and J.~Mandula, ``{All Possible Symmetries of the S Matrix},''
  {\em Phys. Rev.} {\bf 159} (1967)
1251--1256.

\bibitem{Aragone:1979hx}
C.~Aragone and S.~Deser, ``{Consistency Problems of Hypergravity},'' {\em Phys.
  Lett.} {\bf B86} (1979)
161.

\bibitem{Weinberg:1980kq}
S.~Weinberg and E.~Witten, ``{Limits on Massless Particles},'' {\em Phys.
  Lett.} {\bf B96} (1980)
59.

\bibitem{Witten:1988hc}
E.~Witten, ``{(2+1)-Dimensional Gravity as an Exactly Soluble System},'' {\em
  Nucl. Phys.} {\bf B311} (1988)
46.

\bibitem{Ammon:2013hba}
M.~Ammon, A.~Castro, and N.~Iqbal, ``{Wilson Lines and Entanglement Entropy in
  Higher Spin Gravity},'' {\em JHEP} {\bf 10} (2013) 110,
\href{http://www.arXiv.org/abs/1306.4338}{{\tt 1306.4338}}.

\bibitem{deBoer:2013vca}
J.~de~Boer and J.~I. Jottar, ``{Entanglement Entropy and Higher Spin Holography
  in AdS$_3$},'' {\em JHEP} {\bf 04} (2014) 089,
\href{http://www.arXiv.org/abs/1306.4347}{{\tt 1306.4347}}.

\bibitem{Bagchi:2009my}
A.~Bagchi and R.~Gopakumar, ``{Galilean Conformal Algebras and AdS/CFT},'' {\em
  JHEP} {\bf 07} (2009) 037,
\href{http://www.arXiv.org/abs/0902.1385}{{\tt 0902.1385}}.

\bibitem{Bagchi:2010zz}
A.~Bagchi, ``{Correspondence between Asymptotically Flat Spacetimes and
  Nonrelativistic Conformal Field Theories},'' {\em Phys. Rev. Lett.} {\bf 105}
  (2010)
171601.

\bibitem{Barnich:2006av}
G.~Barnich and G.~Compere, ``{Classical central extension for asymptotic
  symmetries at null infinity in three spacetime dimensions},'' {\em Class.
  Quant. Grav.} {\bf 24} (2007) F15--F23,
\href{http://www.arXiv.org/abs/gr-qc/0610130}{{\tt gr-qc/0610130}}.

\bibitem{Bagchi:2012yk}
A.~Bagchi, S.~Detournay, and D.~Grumiller, ``{Flat-Space Chiral Gravity},''
  {\em Phys. Rev. Lett.} {\bf 109} (2012) 151301,
\href{http://www.arXiv.org/abs/1208.1658}{{\tt 1208.1658}}.

\bibitem{Bagchi:2012xr}
A.~Bagchi, S.~Detournay, R.~Fareghbal, and J.~Simón, ``{Holography of 3D Flat
  Cosmological Horizons},'' {\em Phys. Rev. Lett.} {\bf 110} (2013), no.~14,
  141302,
\href{http://www.arXiv.org/abs/1208.4372}{{\tt 1208.4372}}.

\bibitem{Barnich:2012xq}
G.~Barnich, ``{Entropy of three-dimensional asymptotically flat cosmological
  solutions},'' {\em JHEP} {\bf 10} (2012) 095,
\href{http://www.arXiv.org/abs/1208.4371}{{\tt 1208.4371}}.

\bibitem{Barnich:2012aw}
G.~Barnich, A.~Gomberoff, and H.~A. Gonzalez, ``{The Flat limit of three
  dimensional asymptotically anti-de Sitter spacetimes},'' {\em Phys. Rev.}
  {\bf D86} (2012) 024020,
\href{http://www.arXiv.org/abs/1204.3288}{{\tt 1204.3288}}.

\bibitem{Bagchi:2012cy}
A.~Bagchi and R.~Fareghbal, ``{BMS/GCA Redux: Towards Flatspace Holography from
  Non-Relativistic Symmetries},'' {\em JHEP} {\bf 10} (2012) 092,
\href{http://www.arXiv.org/abs/1203.5795}{{\tt 1203.5795}}.

\bibitem{Bagchi:2013bga}
A.~Bagchi, ``{Tensionless Strings and Galilean Conformal Algebra},'' {\em JHEP}
  {\bf 05} (2013) 141,
\href{http://www.arXiv.org/abs/1303.0291}{{\tt 1303.0291}}.

\bibitem{Bagchi:2013lma}
A.~Bagchi, S.~Detournay, D.~Grumiller, and J.~Simon, ``{Cosmic Evolution from
  Phase Transition of Three-Dimensional Flat Space},'' {\em Phys. Rev. Lett.}
  {\bf 111} (2013), no.~18, 181301,
\href{http://www.arXiv.org/abs/1305.2919}{{\tt 1305.2919}}.

\bibitem{Costa:2013vza}
R.~N. Caldeira~Costa, ``{Aspects of the zero $\Lambda$ limit in the AdS/CFT
  correspondence},'' {\em Phys. Rev.} {\bf D90} (2014), no.~10, 104018,
\href{http://www.arXiv.org/abs/1311.7339}{{\tt 1311.7339}}.

\bibitem{Fareghbal:2013ifa}
R.~Fareghbal and A.~Naseh, ``{Flat-Space Energy-Momentum Tensor from BMS/GCA
  Correspondence},'' {\em JHEP} {\bf 03} (2014) 005,
\href{http://www.arXiv.org/abs/1312.2109}{{\tt 1312.2109}}.

\bibitem{Krishnan:2013tza}
C.~Krishnan and S.~Roy, ``{Desingularization of the Milne Universe},'' {\em
  Phys. Lett.} {\bf B734} (2014) 92--95,
\href{http://www.arXiv.org/abs/1311.7315}{{\tt 1311.7315}}.

\bibitem{Krishnan:2013wta}
C.~Krishnan, A.~Raju, and S.~Roy, ``{A Grassmann path from $AdS_3$ to flat
  space},'' {\em JHEP} {\bf 03} (2014) 036,
\href{http://www.arXiv.org/abs/1312.2941}{{\tt 1312.2941}}.

\bibitem{Bagchi:2013qva}
A.~Bagchi and R.~Basu, ``{3D Flat Holography: Entropy and Logarithmic
  Corrections},'' {\em JHEP} {\bf 03} (2014) 020,
\href{http://www.arXiv.org/abs/1312.5748}{{\tt 1312.5748}}.

\bibitem{Detournay:2014fva}
S.~Detournay, D.~Grumiller, F.~Schöller, and J.~Simón, ``{Variational
  principle and one-point functions in three-dimensional flat space Einstein
  gravity},'' {\em Phys. Rev.} {\bf D89} (2014), no.~8, 084061,
\href{http://www.arXiv.org/abs/1402.3687}{{\tt 1402.3687}}.

\bibitem{Barnich:2014kra}
G.~Barnich and B.~Oblak, ``{Notes on the BMS group in three dimensions: I.
  Induced representations},'' {\em JHEP} {\bf 06} (2014) 129,
\href{http://www.arXiv.org/abs/1403.5803}{{\tt 1403.5803}}.

\bibitem{Barnich:2014cwa}
G.~Barnich, L.~Donnay, J.~Matulich, and R.~Troncoso, ``{Asymptotic symmetries
  and dynamics of three-dimensional flat supergravity},'' {\em JHEP} {\bf 08}
  (2014) 071,
\href{http://www.arXiv.org/abs/1407.4275}{{\tt 1407.4275}}.

\bibitem{Riegler:2014bia}
M.~Riegler, ``{Flat space limit of higher-spin Cardy formula},'' {\em Phys.
  Rev.} {\bf D91} (2015), no.~2, 024044,
\href{http://www.arXiv.org/abs/1408.6931}{{\tt 1408.6931}}.

\bibitem{Fareghbal:2014qga}
R.~Fareghbal and A.~Naseh, ``{Aspects of Flat/CCFT Correspondence},'' {\em
  Class. Quant. Grav.} {\bf 32} (2015) 135013,
\href{http://www.arXiv.org/abs/1408.6932}{{\tt 1408.6932}}.

\bibitem{Fareghbal:2014kfa}
R.~Fareghbal and S.~M.~Hosseini, ``{Holography of 3D Asymptotically Flat Black Holes},'' {\em
  Phys. Rev.} {\bf D91} (2015) 084025,
\href{http://www.arXiv.org/abs/1412.2569}{{\tt 1412.2569}}.

\bibitem{Bagchi:2009pe}
A.~Bagchi, R.~Gopakumar, I.~Mandal, and A.~Miwa, ``{GCA in 2d},'' {\em JHEP}
  {\bf 08} (2010) 004,
\href{http://www.arXiv.org/abs/0912.1090}{{\tt 0912.1090}}.

\bibitem{Bagchi:2010vw}
A.~Bagchi, ``{Topologically Massive Gravity and Galilean Conformal Algebra: A
  Study of Correlation Functions},'' {\em JHEP} {\bf 02} (2011) 091,
\href{http://www.arXiv.org/abs/1012.3316}{{\tt 1012.3316}}.

\bibitem{Bagchi:2015wna}
A.~Bagchi, D.~Grumiller, and W.~Merbis, ``{Stress tensor correlators in
  three-dimensional gravity},''
\href{http://www.arXiv.org/abs/1507.05620}{{\tt 1507.05620}}.

\bibitem{Castro:2014tta}
A.~Castro, S.~Detournay, N.~Iqbal, and E.~Perlmutter, ``{Holographic
  entanglement entropy and gravitational anomalies},'' {\em JHEP} {\bf 07}
  (2014) 114,
\href{http://www.arXiv.org/abs/1405.2792}{{\tt 1405.2792}}.

\bibitem{Castro:2014mza}
A.~Castro and E.~Llabrés, ``{Unravelling Holographic Entanglement Entropy in
  Higher Spin Theories},'' {\em JHEP} {\bf 03} (2015) 124,
\href{http://www.arXiv.org/abs/1410.2870}{{\tt 1410.2870}}.

\bibitem{Achucarro:1987vz}
A.~Achucarro and P.~K. Townsend, ``{A Chern-Simons Action for Three-Dimensional
  anti-De Sitter Supergravity Theories},'' {\em Phys. Lett.} {\bf B180} (1986)
89.

\bibitem{Horowitz:1990ap}
G.~T. Horowitz and A.~R. Steif, ``{Singular string solutions with nonsingular
  initial data},'' {\em Phys. Lett.} {\bf B258} (1991)
91--96.

\bibitem{FigueroaO'Farrill:2001nx}
J.~M. Figueroa-O'Farrill and J.~Simon, ``{Generalized supersymmetric
  fluxbranes},'' {\em JHEP} {\bf 12} (2001) 011,
\href{http://www.arXiv.org/abs/hep-th/0110170}{{\tt hep-th/0110170}}.

\bibitem{Liu:2002kb}
H.~Liu, G.~W. Moore, and N.~Seiberg, ``{Strings in time dependent orbifolds},''
  {\em JHEP} {\bf 10} (2002) 031,
\href{http://www.arXiv.org/abs/hep-th/0206182}{{\tt hep-th/0206182}}.

\bibitem{Simon:2002ma}
J.~Simon, ``{The Geometry of null rotation identifications},'' {\em JHEP} {\bf
  06} (2002) 001,
\href{http://www.arXiv.org/abs/hep-th/0203201}{{\tt hep-th/0203201}}.

\bibitem{Cornalba:2002fi}
L.~Cornalba and M.~S. Costa, ``{A New cosmological scenario in string
  theory},'' {\em Phys. Rev.} {\bf D66} (2002) 066001,
\href{http://www.arXiv.org/abs/hep-th/0203031}{{\tt hep-th/0203031}}.

\bibitem{Cornalba:2003kd}
L.~Cornalba and M.~S. Costa, ``{Time dependent orbifolds and string
  cosmology},'' {\em Fortsch. Phys.} {\bf 52} (2004) 145--199,
\href{http://www.arXiv.org/abs/hep-th/0310099}{{\tt hep-th/0310099}}.

\bibitem{Gary:2014ppa}
M.~Gary, D.~Grumiller, M.~Riegler, and J.~Rosseel, ``{Flat space (higher spin)
  gravity with chemical potentials},'' {\em JHEP} {\bf 01} (2015) 152,
\href{http://www.arXiv.org/abs/1411.3728}{{\tt 1411.3728}}.

\bibitem{Hosseini:2015uba}
S.~M. Hosseini and A.~Veliz-Osorio, ``{Gravitational anomalies, entanglement
  entropy, and flat-space holography},''
\href{http://www.arXiv.org/abs/1507.06625}{{\tt 1507.06625}}.

\bibitem{Calabrese:2007rg}
P.~Calabrese and J.~Cardy, ``{Quantum Quenches in Extended Systems},'' {\em J.
  Stat. Mech.} {\bf 0706} (2007) P06008,
\href{http://www.arXiv.org/abs/0704.1880}{{\tt 0704.1880}}.

\bibitem{Calabrese:2012ew}
P.~Calabrese, J.~Cardy, and E.~Tonni, ``{Entanglement negativity in quantum
  field theory},'' {\em Phys. Rev. Lett.} {\bf 109} (2012) 130502,
\href{http://www.arXiv.org/abs/1206.3092}{{\tt 1206.3092}}.

\bibitem{Perlmutter:2015vma}
E.~Perlmutter, M.~Rangamani, and M.~Rota, ``{Central Charges and the Sign of
  Entanglement in 4D Conformal Field Theories},'' {\em Phys. Rev. Lett.} {\bf
  115} (2015), no.~17, 171601,
\href{http://www.arXiv.org/abs/1506.01679}{{\tt 1506.01679}}.

\bibitem{Lewkowycz:2013nqa}
A.~Lewkowycz and J.~Maldacena, ``{Generalized gravitational entropy},'' {\em
  JHEP} {\bf 08} (2013) 090,
\href{http://www.arXiv.org/abs/1304.4926}{{\tt 1304.4926}}.

\end{thebibliography}
\end{document}